\documentclass[twocolumn,aps,prl,superscriptaddress]{revtex4-2}

\usepackage{graphicx}
\usepackage{xcolor}

\usepackage[T1]{fontenc}
\usepackage[utf8]{inputenc}
\usepackage{graphicx}
\usepackage{epsfig}
\usepackage{adjustbox}
\usepackage{amssymb,amsmath,amsopn,amsfonts}
\usepackage{colordvi}
\usepackage{color}
\usepackage{subeqnarray}
\usepackage{booktabs,tabularx}
\usepackage{siunitx}
\usepackage{bbold}
\usepackage{xcolor}
\usepackage{siunitx}
\usepackage{braket}
\usepackage{bbold}
\usepackage{mathtools}
\usepackage{soul}
\colorlet{blue}{black}
\usepackage{comment}
\usepackage{ragged2e} % gives \justifying
\usepackage{caption}
\usepackage{subcaption}
\DeclareCaptionJustification{just}{\justifying}
\usepackage{float}
\renewcommand{\thesection}{\arabic{section}}
\renewcommand{\thesubsection}{\arabic{section}.\arabic{subsection}}
\makeatletter
\def\p@subsection{}
\makeatother

\makeatletter
\def\NAT@def@citea{\def\@citea{\NAT@separator}}
\makeatother

\usepackage[hidelinks]{hyperref}

\begin{document}

\title{Heralded generation of a three-mode NOON state}

\author{Sukhjit P. Singh}
\affiliation{Queensland Quantum and Advanced Technologies Research Institute, Centre for Quantum Computation and Communication Technology, Griffith  University, Yuggera Country,   Brisbane,  Queensland,  4111  Australia}

\author{Elnaz Bazzazi}
\affiliation{ Department of Physics, Humboldt University of Berlin, Berlin, 12489 Germany}

\author{Diego N. Bernal-Garc\'{\i}a}
\email{dn.bernalgarcia@gmail.com}
\affiliation{Queensland Quantum and Advanced Technologies Research Institute, Centre for Quantum Computation and Communication Technology, Griffith  University, Yuggera Country,   Brisbane,  Queensland,  4111  Australia}
\author{Simon White}
\affiliation{Queensland Quantum and Advanced Technologies Research Institute, Centre for Quantum Computation and Communication Technology, Griffith  University, Yuggera Country,   Brisbane,  Queensland,  4111  Australia}
\author{Hassan Jamal Latief}
\affiliation{Centre for Quantum Computation and Communication Technology, School of Physics, The University of New South Wales, Sydney, NSW 2052, Australia}
\author{Alison Goldingay}
\affiliation{Centre for Quantum Computation and Communication Technology, School of Physics, The University of New South Wales, Sydney, NSW 2052, Australia}
\author{Sven Rogge}
\affiliation{Centre for Quantum Computation and Communication Technology, School of Physics, The University of New South Wales, Sydney, NSW 2052, Australia}
\author{Sergei Slussarenko}
\affiliation{Queensland Quantum and Advanced Technologies Research Institute, Centre for Quantum Computation and Communication Technology, Griffith  University, Yuggera Country,   Brisbane,  Queensland,  4111  Australia}
\author{Farzad Ghafari}
\email{f.ghafari@griffith.edu.au}
\affiliation{Queensland Quantum and Advanced Technologies Research Institute, Centre for Quantum Computation and Communication Technology, Griffith  University, Yuggera Country, Brisbane,  Queensland,  4111  Australia}
\author{Emanuele Polino}
\email{e.polino@griffith.edu.au}
\affiliation{Queensland Quantum and Advanced Technologies Research Institute, Centre for Quantum Computation and Communication Technology, Griffith  University, Yuggera Country,   Brisbane,  Queensland,  4111  Australia}
\author{Nora Tischler}
\email{n.tischler@griffith.edu.au}
\affiliation{Queensland Quantum and Advanced Technologies Research Institute, Centre for Quantum Computation and Communication Technology, Griffith  University, Yuggera Country,   Brisbane,  Queensland,  4111  Australia}

\date{\today}
\begin{abstract}
 Entangled states of photons form the foundation of quantum communication, computation, and metrology. Yet their generation remains fundamentally constrained: in the absence of intrinsic photon–photon interactions, the generation of such states is inherently probabilistic rather than deterministic. The prevalent technique of post-selection verifies the creation of an entangled state by detecting and thus destroying it. Heralding offers a solution in which measuring ancillary photons in auxiliary modes signals the state generation without the need to measure it. Here, we \textcolor{blue}{introduce and experimentally demonstrate a scheme} to generate a three-mode NOON state, where the detection of a single photon in one heralding mode signifies the presence of the state in three target modes. We validate the generated state by estimating a fidelity of $0.823\pm0.018$ with respect to an ideal three-mode NOON state and certifying genuine multipartite entanglement. By virtue of the high success probability and small resource overhead of our scheme, our work provides a theoretical and experimental stepping stone for entangled multi-mode state generation, which is realizable with current technology. These multi-mode entangled states represent a key direction for linear optical quantum information that is complementary to multi-qubit state encoding.
 
\end{abstract}

\maketitle

\paragraph*{Introduction---}
Entanglement is a defining feature of quantum mechanics, marking a clear departure from classical behaviour. Beyond their foundational significance, entangled states of photons also serve as a key resource for achieving quantum advantage in computing, communication, and metrology~\cite{Prevedel, Panreview, Flamini_2019}. However, the generation of these photonic states remains generally a probabilistic process owing to negligible photon-photon interactions~\cite{O'Brien2009, walther}. This leads to the fundamental challenge of determining when the desired state has been successfully generated. Most protocols for verifying state generation rely on post-selection, a destructive approach that measures all outputs to certify the produced state. Heralding overcomes this limitation by detecting ancillary photons in auxiliary modes, thereby providing an independent signal that flags the creation of the entangled state without disturbing it~\cite{bartolucci2021creationentangledphotonicstates}. This capability is essential for fusion-based quantum computing~\cite{bartolucci2023fusion}, and for quantum communication, in which the heralding signal provides loss tolerance~\cite{luswapping}. Heralding will likewise strengthen quantum metrology by optimizing the use of resources~\cite{ourreview}. 

The heralded generation of maximally entangled states is an enduring research field that has recently been reignited. Notable experiments include the heralded generation of Bell states~\cite{barz_heralded_2010, pittman2003, Sfiliwa2003Conditional,wagenknecht_experimental_2010,HeraldedBell,RussianHeraldedBell}, two-mode NOON states~\cite{Eisenberg2005, kim2009, Smith2008,  Ra2015, Vergyris2016, jonathan, Mitchell2004} and most recently, the GHZ state~\cite{PanGHZ, WaltherGHZ, quandelaGHZ}.
Two-mode NOON states, $(\ket{N,0} + \ket{0, N})/\sqrt{2}$, came to prominence as the gold standard for single-phase sensing in the absence of losses~\cite{dowling, slussarenko2017unconditional}. 
These states can be extended to multiple modes, where they again serve as a resource for important tasks in quantum metrology.  
For example, multi-mode NOON states are useful for multiphase estimation~\cite{hong,hong2,namkung2024optimal} and distributed sensing ~\cite{distributedsensing2025}, with potential applications ranging from quantum imaging and global clock synchronization to sensor networks and quantum algorithms~\cite{Zhang_2021, Barbieri2024, Gebhart2021}.

Three-mode NOON states are maximally entangled states, defined as a coherent superposition of $N$ photons in one mode and none in the other two modes: 
    $|\psi_3^{N}\rangle = (\ket{N,0,0} + e^{i\alpha_1}\ket{0, N,0} + e^{i\alpha_2}\ket{0,0, N})/\sqrt{3}$, 
where $\alpha_1$ and $\alpha_2$ are relative phases. 
Although proposals exist for generating three-mode NOON states spanning photons~\cite{ZHANGlowprobNOON2019,Zhang2017}, atoms~\cite{ultracoldNOON, atomNOON}, and solitons~\cite{solitonNOON}, experimental realization remains extremely challenging: To date, a three-mode NOON state has only been reported in one experiment, and relied entirely on post-selection \cite{hong2}. However, to allow for unrestricted usage, these states must be generated in a heralded fashion~\cite{ourreview}.

Here, we report the experimental generation of a three-mode NOON state, 
\begin{equation}
    \ket{\psi_3^2} = \frac{1}{\sqrt{3}}\left(\ket{2,0,0} + e^{i\alpha_1}\ket{0,2,0} + e^{i\alpha_2}\ket{0,0,2}\right),
    \label{eq:realNOON}
\end{equation}
whose successful creation is heralded by the detection of a single photon in an auxiliary mode. 
For this purpose, we theoretically introduce and experimentally demonstrate a scheme utilizing three single photons distributed across four modes. 
Our protocol employs an interferometric module, which acts jointly on the path and polarization degrees of freedom~\cite{Englertgate, Guanggate2020} and is optimized for simplicity and probability of success. The scheme features a nominal success probability of $0.25$.
We then implement this protocol in a proof-of-principle photonic experiment and generate the heralded three-mode NOON state of Eq.~\eqref{eq:realNOON}. We achieve a success probability of $0.237 \pm 0.009$ and verify the state creation with a fidelity of $F = \bra{\psi_{3}^2} \rho \ket{\psi_{3}^2} = 0.823 \pm 0.018$, where $\rho$ denotes the experimentally prepared state. This verification relies on a tailored method that avoids challenges of full quantum state tomography. Moreover, the achieved fidelity surpasses the threshold for genuine multipartite entanglement by more than eight standard deviations. % In addition to the scheme for three-mode NOON states, we propose an extension for generating heralded two-photon NOON states in an arbitrary number of modes.

Achieving high success probability and modest resource requirements, our work demonstrates a decisive advance towards practical schemes for heralded multi-mode entangled states generation, fundamental for linear optical quantum information. In addition to the scheme for two-photon three-mode NOON states, we propose \textcolor{blue}{generalizations to reach arbitrary numbers of modes and photons.}

\paragraph*{Unitary transformation for heralded state generation---}    

\begin{figure}[ht]
    \centering
    \includegraphics[width=1.0\columnwidth]{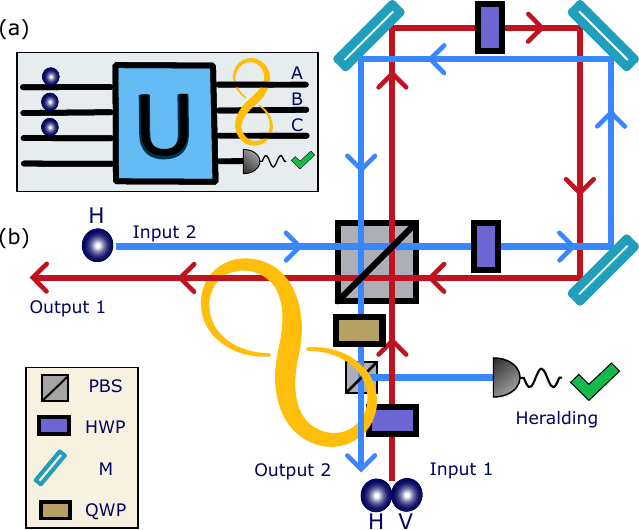}
    \caption{ \textbf{Heralded generation of a three-mode NOON state via linear optics}: (a) From an input of three single photons, a four-mode unitary transformation generates the desired coherent superposition of two photons in three modes (A,B,C) upon detecting a photon in the fourth mode. (b) The experimental setup for state generation consists of a displaced Sagnac interferometer constructed using a polarizing beam splitter and wave plates. The heralding process is schematically indicated, highlighting the projection onto the three-mode NOON state in the target modes upon detection of a photon in the heralding mode. The interferometer provides a stable and flexible platform for realizing four-mode unitary transformations~\cite{Englertgate, Guanggate2020}. QWP: quarter-wave plate; HWP: half-wave plate; M: mirror; PBS: polarizing beam splitter. The red and blue lines with arrows indicate the two spatial paths of the interferometer.}
    \label{fig: NOONOON}
\end{figure}

We perform a four-mode unitary transformation on three single photons as illustrated by Fig.~\ref{fig: NOONOON}(a). We generate the desired two-photon three-mode state with a nominal success probability $|\gamma|^{2} = 0.25$, heralded by the detection of one photon in the auxiliary mode:

\begin{align}
    U\ket{1,1,1,0}_{\mathrm{input}} 
    = \gamma\ket{\psi_3^2}_{\text{target}}\ket{1}_{\mathrm{herald}} + \dots \;, 
\end{align}
where the omitted terms correspond to the absence of the heralding signal and are therefore discarded.
We use a combination of degrees of freedom to encode the four modes spanned by the two orthogonal polarizations and two path modes.
We define the mode convention through the following single-photon states:
\begin{equation}
\begin{split}
    a^\dagger_{\mathrm{H},\,\mathrm{path}1} \ket{\mathbf{0}} &= \ket{100}_{\text{target}}\ket{0}_{\text{herald}}, \\
    a^\dagger_{\mathrm{V},\,\mathrm{path}1} \ket{\mathbf{0}} &= \ket{010}_{\text{target}}\ket{0}_{\text{herald}}, \\
    a^\dagger_{\mathrm{H},\,\mathrm{path}2} \ket{\mathbf{0}} &= \ket{001}_{\text{target}}\ket{0}_{\text{herald}}, \\
    a^\dagger_{\mathrm{V},\,\mathrm{path}2} \ket{\mathbf{0}} &= \ket{000}_{\text{target}}\ket{1}_{\text{herald}} \;,
\end{split}
\end{equation}
where H and V denote horizontal and vertical polarizations, and the three modes of the target state are labeled $A$, $B$, and $C$ in Fig.~\ref{fig: NOONOON}(a). Note that, if desired, the target state modes could be deterministically converted into purely path encoding.

We employ a gradient–descent search in the space of four-mode unitary transformations to identify unitaries that maximize the fidelity 
of the output state with respect to Eq.~\eqref{eq:realNOON}, given the separable input state $\ket{1,1,1,0}$. Among the multiple solutions obtained with unit fidelity and success probability of $0.25$, we select the unitary with the simplest experimental implementation based on the number of optical elements.
The chosen transformation [see Sec.~S1 of the Supplemental Material (SM)] is implemented by the setup shown in Fig.~\ref{fig: NOONOON}(b), a polarization-based displaced Sagnac interferometer.

\paragraph{Measurement protocol---} Full quantum state tomography of the three-mode two-photon state is experimentally challenging, owing to the difficulty of characterizing coherent superpositions of multiphoton and vacuum components across several modes. 
Instead, we employ a targeted measurement protocol that exploits the structure of the three-mode NOON state in Eq.~\eqref{eq:realNOON} to extract the state fidelity without complete tomographic reconstruction.
 
The measurement strategy relies on two types of measurements, all of which are conditioned on the heralding signal: a) projection onto the three-mode Fock states with two photons in total (i.e., populations); and b) probing information about coherences. 

The coherence measurements rely on vacuum projection of individual modes $k$, reducing the heralded three-mode state $\rho$ to the conditional states $\rho^{(ij)}$  within two-mode two-photon subspaces. Here $(i,j,k)$ denotes any permutation of $(A,B,C)$, with $k$ projected onto vacuum and $(i,j)$ labeling the remaining modes.%Here $(i, j, k) \in \{A, B, C\}$ and are all distinct from each other. 
When mode $k$ is projected onto vacuum, the remaining modes $i$ and $j$ are described by a conditional unnormalized 
density matrix $\rho^{(ij)} = \prescript{}{k}{\bra{0}} \rho \ket{0}_k$. For instance, when $k=C$,
$\rho^{(AB)}$ exists in the subspace spanned by $\{\ket{2,0}_{AB}, \ket{0,2}_{AB}, \ket{1,1}_{AB}\}$ (see Fig.~S2 of the SM). 
To extract the coherence between the $\ket{2,0}_{ij}$ and $\ket{0,2}_{ij}$ components, we perform a two-mode interferometric measurement consisting of a wave-plate transformation $U_{\mathrm{meas}}(\theta) = \text{HWP}(\theta)\,\text{QWP}(\pi/4)$ followed by a polarizing beam splitter (PBS).
Photon detectors at the PBS output ports record coincidence events, and the measured coincidence fringe $C_{ij}$ as a function of the HWP angle $\theta$ is
\begin{equation}
C_{ij}(\theta) = \frac{A_{ij}}{2} + \frac{V_{ij}}{2} \cos(8\theta + \varphi_{ij}).
\label{eq:coincidence}
\end{equation}
Here, the offset $A_{ij}$, visibility $V_{ij}$, and phase $\varphi_{ij}$ are directly related to the elements of the prepared density matrix $\rho$ (see S2.2 of the SM for the explicit derivation).
In particular, the visibility $V_{ij}$ quantifies the magnitude of coherence between the $\ket{2,0}_{i,j}$ and $\ket{0,2}_{i,j}$ basis states, while $\varphi_{ij}$ represents its complex phase. 
Through the extraction of both $V_{ij}$ and $\varphi_{ij}$ from the sinusoidal fit, this measurement protocol enables complete reconstruction of the complex off-diagonal elements $\rho_{020,002}$, $\rho_{200,002}$, and $\rho_{200,020}$.
 
By repeating this procedure for all three mode combinations, projecting modes $A$, $B$, and $C$ onto the vacuum in turn, we obtain three independent coincidence fringes $C_{BC}(\theta)$, $C_{AC}(\theta)$, and $C_{AB}(\theta)$.
The coherence measurements alone allow for the evaluation of lower and upper bounds on the fidelity with respect to the three-mode NOON state of Eq.~\eqref{eq:realNOON}. Combined with the population measurements, the coherence measurements provide sufficient information to estimate the fidelity.
Further details about the fidelity estimation are provided in Sec.~S2 of the SM.

This measurement strategy enables robust state characterization without requiring full tomographic reconstruction.

\paragraph*{Experimental implementation---}
The input state $\ket{1,1,1,0}$ is prepared by pumping a periodically poled potassium titanyl phosphate (ppKTP) crystal with a 775~nm pulsed picosecond laser. The crystal is pumped in two different locations to generate two pairs of orthogonally polarized photons via two simultaneous type-II spontaneous parametric down-conversion events. More details about the source are provided in the SM, see S3. Three of the four photons are sent into the state-generation setup (Fig.~\ref{fig: NOONOON}), while the fourth is used as a trigger.
To overcome the non-deterministic preparation of the input state, $\ket{1,1,1,0}$, 
we certify the correct input state by considering four-fold coincidence events in the final measurements.

\begin{figure*}[t]
    \centering
    \includegraphics[width=0.8\linewidth]{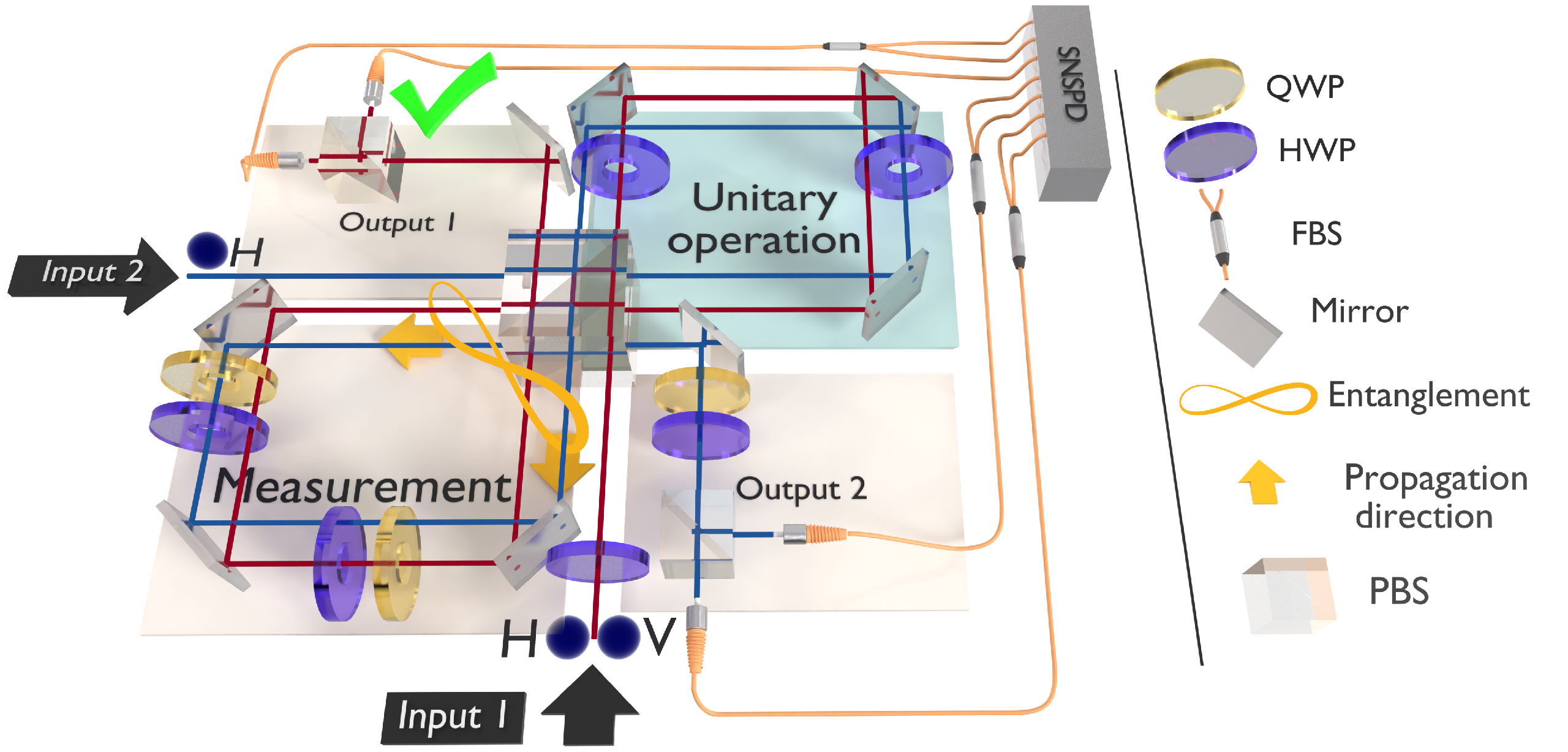}
    \caption{\textbf{Interferometric scheme for heralded NOON state generation and characterization}. State preparation relies upon the HWP in input 1 and the first interferometer (unitary operation panel), followed by a QWP in the blue path in the subsequent interferometer. 
   The measurement stage is realized using the second interferometer (measurement panel) and polarization projection in both outputs.
   The wave plates inside the interferometers have a hole in the center such that the polarization of only one path is transformed. Detection is performed using SNSPDs, with fiber beam splitters enabling pseudo-photon-number-resolving measurements on all three target modes. QWP: quarter-wave plate; HWP: half-wave plate; PBS: polarizing beam splitter; FBS: fiber beam splitter.}
    \label{fig:elnaz}
\end{figure*}

For both types of measurements, photon detection is performed using superconducting nanowire single-photon detectors (SNSPDs). Pseudo–photon number resolution is enabled by fanning out each target mode into two using a fiber beam splitter before detection~\cite{ourreview}. To implement the coherence measurements discussed above, we use a second displaced Sagnac interferometer, shown in the measurement panel of Fig.~\ref{fig:elnaz}, followed by a polarization measurement.
Coincidence probabilities as a function of $\theta$ are evaluated from coincidence counts. Thereafter, the required coherences are extracted from the data, and both an estimate of the fidelity and its lower and upper bounds with respect to Eq.~\eqref{eq:realNOON} are determined.

\paragraph*{Experimental results---}
 
\begin{figure*}[t]
    \centering
    \includegraphics[width=\linewidth]{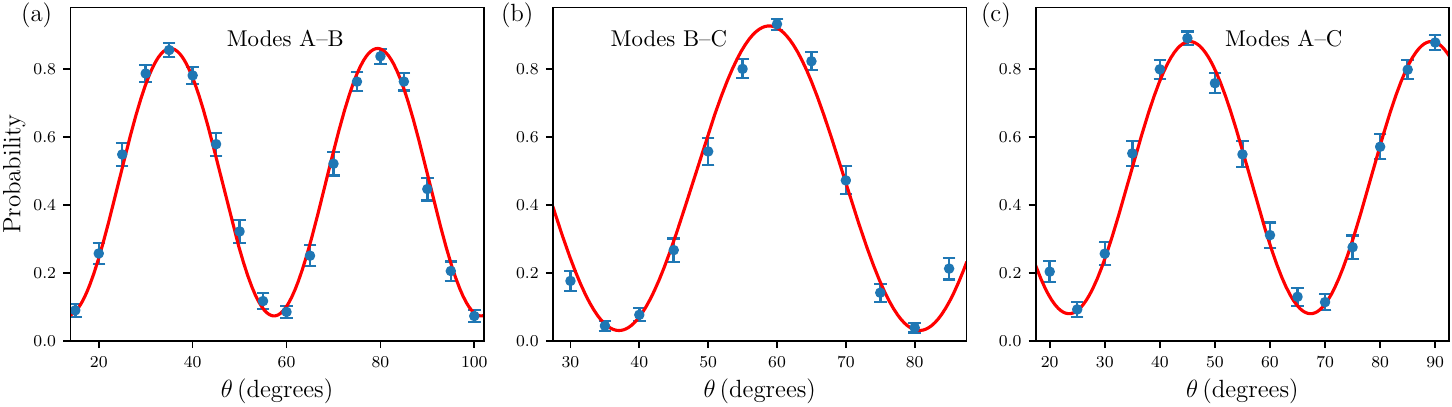}
    \caption{The heralded three-mode state is projected onto a two-photon two-mode subspace by conditioning on vacuum in the remaining mode. To probe the coherence of the state in this subspace, we interfere the two modes after mapping them into the polarization basis of a single spatial mode.   
    The resulting coincidence probabilities are shown as a function of the measurement half-wave plate angle $\theta$ for pairwise combinations of modes A and B 
    (a), B and C  (b), and A and C (c). Each probability data point (blue points) is obtained from an average of 1039 four-fold coincidence counts accumulated over 1800 s. The resulting data are fitted with the function given in Eq.~\eqref{eq:coincidence} (red curve), and the uncertainties in the measured counts represent Poissonian statistics.}
    \label{fig: Oscillation}
\end{figure*}

Following the measurement protocol described above and further detailed in S2.2 of the SM, we evaluate the fidelity of the generated three-mode NOON state and find $F = 0.823 \pm 0.018$ with respect to Eq.~\eqref{eq:realNOON}, where $\alpha_{1} = 1.568 \pm 0.030 $ and $\alpha_{2} = -0.262 \pm 0.033$ (in radians). This fidelity is compatible with the lower and upper bounds $F \in [0.817, 0.836]$, evaluated using only the coherence measurements shown in Fig.~\ref{fig: Oscillation} (for more detail, see S2.3 of the SM). %We estimate the experimental success probability of our scheme to be $|\gamma_{\mathrm{exp}}|^2=0.237\pm0.009$. 
Conditioned on the availability of the input state 
$\ket{1,1,1,0}$, whose event rate depends on the choice of single-photon source (see S3.3 of the SM), the experimental success probability of our scheme is estimated to be $|\gamma_{\mathrm{exp}}|^2=0.237\pm0.009$. This is obtained by dividing the four-fold events in which heralding was successful using the unitary transformation, by the four-folds while implementing an identity transformation. 

Furthermore, we study the presence of genuine multipartite entanglement (GME). Three-mode states exhibiting GME cannot be decomposed into a statistical mixture of states that are separable over any bipartition of the modes, i.e., they are not biseparable~\cite{GUHNE2009, Palazuelos2022}.  We certify GME by comparing the estimated fidelity with respect to Eq.~\eqref{eq:realNOON}, against the largest fidelity $F_\mathrm{bs}$ attainable with biseparable states. 

Following the procedure described in the Appendix of Ref.~\cite{Bourennane2004}, we determine  $F_\mathrm{bs}$  by considering all bipartitions of the target state in Eq.~\eqref{eq:realNOON}.
For each bipartition, we compute the Schmidt decomposition and identify the largest squared Schmidt coefficient.
The maximum of these values across all bipartitions yields the GME threshold, $F_\mathrm{bs} = 2/3$. 
The full derivation is provided in
Sec.~S2.4 of the SM.   
Our estimated fidelity %$F = 0.823 \pm 0.018$ 
exceeds this threshold by more than 8 standard deviations, unambiguously certifying genuine tripartite entanglement distributed across all three target modes.

\paragraph*{Discussion---} 
In this work, we demonstrate the heralded generation of a three-mode two-photon NOON state. To this end, we identify a suitable linear optical unitary for state generation providing unit fidelity, success probability of 0.25, and a resource-efficient implementation.  
Our scheme has a higher success probability than previous schemes~\cite{ZHANGlowprobNOON2019, Zhang2017}, while reducing the number of input photons. We implement this scheme in a photonic experiment, where we rigorously characterize the state fidelity and certify genuine three-mode entanglement with high statistical significance. Moreover, in Sec.~S4 of the SM, we provide a generalization of the scheme to
generate two-photon multi-mode NOON states with an arbitrary number of modes.  \textcolor{blue}{Such two-photon multi-mode NOON states have been identified as an important and experimentally relevant class of states for quantum-enhanced multiphase estimation in lossy environments~\cite{namkung2024optimal}.} Remarkably, the resource scaling of \textcolor{blue}{our} generalization is linear in terms of modes and photons. \textcolor{blue}{The scheme can, in principle, be further generalized to arbitrary photon numbers, as discussed in Sec.~5 of the SM.}

Given the correct three-photon input state, the heralded nature of the protocol allows for unrestricted use of the generated state upon detection of only one photon in the heralding mode. In our experiment, the three photons incident on the circuit are generated probabilistically through two simultaneous SPDC processes, and their availability is ensured by four-fold coincidence detection at the output.
This choice of final photon counting is not a fundamental requirement: Using three SPDC sources, one could guarantee the correct three-photon input state is available, where the detection of three single photons heralds their respective partners. Alternatively, deterministic sources with unit collection and multiplexing efficiency could also provide such a guaranteed input state. The photon counting we employ for verification of the input state has the feature of mitigating the effect of losses, which are present in all practical scenarios and are independent of the choice of photon source. 
Importantly, this photon counting, based on the total number of photons across all the modes, allows for the usage of the state.
By contrast, standard post-selection, which occurs based on the number of photons in specific subsets of modes, imposes restrictions on subsequent interference of modes, severely limiting the use of post-selected states~\cite{ourreview}.

We have demonstrated a practical approach to generate a multi-mode NOON state from three single photons, using a bulk-optical setup that can be implemented in standard quantum optics laboratories. This opens a pathway to multi-mode photonic resources beyond post-selected demonstrations. An interesting future application of our heralded state would be multi-phase sensing~\cite{Barbieri2024}.
 
Three-mode NOON states exhibit an almost unit theoretical fidelity ($|\langle \psi_{\mathrm{opt}}^N|\psi_3^N\rangle|^2 \approx \textcolor{blue}{0.993}$)  with the state $|\psi_{\mathrm{opt}}^N\rangle$ identified in Ref.~\cite{barbieri}, which is optimal for simultaneous multi-phase estimation.
In our experiment, the experimentally generated state attains a fidelity of $\langle \psi_{\mathrm{opt}}^2|\, \rho\, |\psi_{\mathrm{opt}}^2\rangle = 0.836 \pm 0.019$ with respect to this optimal target state.

Overall, our work represents an important experimental and theoretical advance toward the generation of increasingly complex heralded photonic states with more modes and photons~\cite{chin2024exponentially, Bhatti_2025, Zhang2017, Takeuchi2023}. 
This approach can be further extended through integrated photonics, providing a scalable and compact platform for implementing complex quantum circuits~\cite{Wang2025} (see Sections S1.2 and S4.3 of the SM). 
%\textcolor{red}{As a}
Through this demonstration of heralded three-mode entangled state generation and the proposal of a general scheme for higher-mode NOON states, %As a demonstration of heralded multi-mode entangled state generation,
our work will contribute to future advances in quantum communication~\cite{distributedsensing2025}, multiphase sensing and quantum algorithms~\cite{barbieri, Gebhart2021}. 

\section{Acknowledgements}
We acknowledge helpful discussions with Howard M. Wiseman. This work was supported by the Australian Research Council; N.T.~is a recipient of an Australian Research Council Discovery Early Career Researcher Award (DE220101082); S.S.~is a recipient of an Australian Research Council Future Fellowship (FT240100352); E.P.~is a recipient of an Australian Research Council Discovery Early Career Researcher Award (DE250100762); the work was in part supported by ARC Grant No.~CE170100012. S.P.S.~acknowledges support from the Australian Government Research Training Program (RTP). F.G.~is supported partly by the Griffith University Postdoctoral Fellowship (GUPF\#58938). E.B.~acknowledges support by BMFTR (project 13N16105).  This material is based upon work supported by the Air Force Office of Scientific Research under Award No.~FA2386-23-1-4086.
%-------------------------------------%
%\bibliography{revbib}

\clearpage
\onecolumngrid

% --- Begin Supplementary Material ---
\appendix
% \documentclass[11pt,a4paper]{article}
%\documentclass[twocolumn,aps,pra,superscriptaddress]{revtex4-2}
% \usepackage{xcolor}
% \usepackage{titlesec}
% \usepackage{xcolor}

% \newenvironment{bluesubsection}[1]
% {
% \subsection{#1}
% \color{blue}
% }
% {
% \color{black}
% }
% \usepackage{amsmath}
% \usepackage{xcolor}

\newenvironment{bluesection}[1]
{
\section{#1}
\color{blue}
}
{
\color{black}
}

%\input{config}

% Section numbering
\renewcommand{\thesection}{S\arabic{section}}
\renewcommand{\thesubsection}{S\arabic{section}.\arabic{subsection}}

% Reset counters for supplemental material
\setcounter{equation}{0}
\setcounter{figure}{0}
\setcounter{table}{0}
\setcounter{page}{1}
\setcounter{section}{0}

% Redefine counters for supplemental material
\renewcommand{\theequation}{S\arabic{equation}}
\renewcommand{\thefigure}{S\arabic{figure}}
\renewcommand{\thetable}{S\arabic{table}}

% For citations in text
\renewcommand{\citenumfont}[1]{S#1}
\renewcommand{\bibnumfmt}[1]{[S#1]}

%\makeatletter
%\renewcommand{\tagform@}[1]{%
%  \begingroup
%  \edef\temp{#1}%
%  \ifnum\pdfstrcmp{\temp}{S10}<0
    %\maketag@@@{\normalfont(#1)}%
  %\else
%     \maketag@@@{\textcolor{blue}{\normalfont(#1)}}%
%   \fi
%   \endgroup
% }
% \makeatother

\onecolumngrid
\thispagestyle{empty}

\begin{center}
\textbf{\large Supplemental Material: Heralded generation of a three-mode NOON state}
\end{center}

\begin{center}
Sukhjit P. Singh,$^{1}$ Elnaz Bazzazi,$^2$ Diego N. Bernal-García,$^{1}$, Simon White,$^{1}$ Alison Goldingay,$^{3}$ \\ Hassan Jamal Latief,$^{3}$ Sven Rogge,$^{3}$ 
Sergei Slussarenko,$^{1}$ Farzad Ghafari,$^{1}$ Emanuele Polino,$^{1}$ and  Nora Tischler$^1$
\end{center}
\vspace{-0.5cm}
\begin{center}
\text{\small \it $^{1}$ Queensland Quantum and Advanced Technologies Research Institute,} \\
\text{\small \it Centre for Quantum Computation and Communication Technology,} \\
\text{\small \it Griffith University, Yuggera Country, Brisbane, Queensland, 4111 Australia} 

\text{\small \it $^{2}$ Department of Physics, Humboldt University of Berlin, Berlin, 12489 Germany}
\text{\small \it $^{3}$ Centre for Quantum Computation and Communication Technology,
School of Physics,}\\
\text{\small \it The University of New South Wales, Sydney, NSW 2052, Australia}
\end{center}
\vspace{-0.5cm}
\begin{center}
    \text{ \small (Dated: \today) }
\end{center}

%-------------------------------------------------------------------------------

\section{S1. Unitary transformation}
\label{supp:unitary_transformation}

The four–mode space relevant for our optical transformation is spanned by the
hybrid path--polarization modes
\[
\{\ket{H,1},\,\ket{V,1},\,\ket{H,2},\,\ket{V,2}\},
\]
where $\ket{P,j}$ denotes a single photon with polarization $P\in\{H,V\}$ in
spatial path $j\in\{1,2\}$.  In this ordered basis,
\begin{equation}
\ket{H,1}=\begin{pmatrix}1\\0\\0\\0\end{pmatrix},\qquad
\ket{V,1}=\begin{pmatrix}0\\1\\0\\0\end{pmatrix},\qquad
\ket{H,2}=\begin{pmatrix}0\\0\\1\\0\end{pmatrix},\qquad
\ket{V,2}=\begin{pmatrix}0\\0\\0\\1\end{pmatrix}.
\end{equation}

For completeness, we recall the correspondence between these polarization–path
modes and the occupation–number notation used in the main text.  Acting on the
vacuum $\ket{\mathbf{0}}$,
\begin{equation}
\begin{split}
a^\dagger_{\mathrm{H},1}\ket{\mathbf{0}} &= \ket{100}_{\mathrm{target}}\ket{0}_{\mathrm{herald}}
  \equiv \ket{H,1},\\
a^\dagger_{\mathrm{V},1}\ket{\mathbf{0}} &= \ket{010}_{\mathrm{target}}\ket{0}_{\mathrm{herald}}
  \equiv \ket{V,1},\\
a^\dagger_{\mathrm{H},2}\ket{\mathbf{0}} &= \ket{001}_{\mathrm{target}}\ket{0}_{\mathrm{herald}}
  \equiv \ket{H,2},\\
a^\dagger_{\mathrm{V},2}\ket{\mathbf{0}} &= \ket{000}_{\mathrm{target}}\ket{1}_{\mathrm{herald}}
  \equiv \ket{V,2}.
\end{split}
\end{equation}

The four–mode unitary to generate a heralded three-mode NOON state from a separable input of three indistinguishable single photons, which is found through numerical optimization, can then be expressed in this basis as:
\begin{align}\label{matrix}
{\color{blue}
U=
\renewcommand{\arraystretch}{1.2}
\begin{pmatrix}
-\frac{1}{\sqrt{2}} & -\frac{1}{\sqrt{2}} & 0 & 0 \\[1mm]
\frac{1}{\sqrt{6}} & -\frac{1}{\sqrt{6}} & -\sqrt{\frac{2}{3}} & 0 \\[1mm]
-\frac{i}{\sqrt{6}} & \frac{i}{\sqrt{6}} & -\frac{i}{\sqrt{6}} & -\frac{1}{\sqrt{2}} \\[1mm]
\frac{1}{\sqrt{6}} & -\frac{1}{\sqrt{6}} & \frac{1}{\sqrt{6}} & \frac{i}{\sqrt{2}}
\end{pmatrix}
}
\end{align}
\subsection{S1.1. Bulk optical implementation}
\begin{figure}[ht]
        \centering
        \includegraphics[width=0.4\columnwidth]{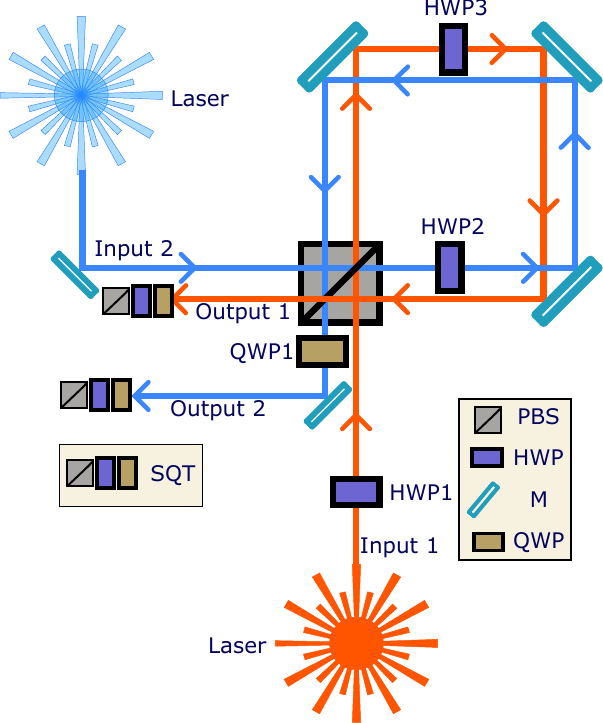}
        \caption{ Simplified schematic for the characterization of the setup with classical light. QWP: quarter-wave plate; HWP: half-wave plate; M: mirror; PBS: polarizing beam splitter; SQT: single-qubit tomography setup. The orange and blue lines with arrows indicate the two spatial paths of the interferometer.}
        \label{fig: NOONOON supplementary}
\end{figure}

The experimental optical realization of this unitary transformation is illustrated in Fig.~\ref{fig: NOONOON supplementary}, where 
\[
U = U_{\mathrm{QWP1}}\,U_{\mathrm{PBS}}\,U_{\mathrm{Mirror}}\,U_{\mathrm{HWP3}}\,U_{\mathrm{Mirror}}\,U_{\mathrm{Mirror}}\,U_{\mathrm{HWP2}}\,U_{\mathrm{PBS}}U_{\mathrm{HWP1}}.
\]
Each wave plate implements a transformation that acts only on the spatial mode in which it is physically placed. Consequently, its action on the two-path, two-polarization Hilbert space is represented by a $4\times4$ block-diagonal unitary,
\[
U_{\mathrm{WP}} =
\begin{pmatrix}
V_{2\times2} & 0 \\
0 & I_{2\times2}
\end{pmatrix}
\quad\text{or}\quad
U_{\mathrm{WP}} =
\begin{pmatrix}
I_{2\times2} & 0 \\
0 & V_{2\times2}
\end{pmatrix},
\]
depending on whether the wave plate is located in the spatial path for input 1 or input 2, respectively. Here $V_{2\times2}$ is the polarization transformation applied to the affected path, while $I_{2\times2}$ denotes the identity acting on the unaffected path. 
The single-path polarization transformations $V_{2\times2}$ corresponding to a half-wave plate (HWP) and a quarter-wave plate (QWP) oriented at an angle $\theta$ are (see Ref.~\cite{SMEnglertgate}):
\[
V_{\mathrm{HWP}}(\theta)
= -i
\begin{pmatrix}
\cos(2\theta) & \sin(2\theta) \\
\sin(2\theta) & -\cos(2\theta)
\end{pmatrix},
\]
\[
V_{\mathrm{QWP}}(\theta)
= \frac{1}{\sqrt{2}}
\begin{pmatrix}
1- i\cos(2\theta) 
    & -i\sin(2\theta) \\
-i\sin(2\theta)
    & 1+ i\cos(2\theta)
\end{pmatrix}.
\]
In the configuration used here, the wave plate angles are set as follows: $\mathrm{HWP1} = \pi/8$, $\mathrm{HWP2} = 0.848\pi$, $\mathrm{HWP3} = 0$, and $\mathrm{QWP1} = 3\pi/4$.
\medskip

The polarizing beam splitter (PBS) separates horizontal and vertical polarizations into distinct spatial paths. In our convention, horizontally polarized photons are transmitted, while vertically polarized photons are routed to the opposite spatial mode and acquire a phase factor of $i$. In particular, a vertically polarized photon entering path~1 exits in path~2 with an additional phase $i$, and vice versa. 
Its action in the four-mode space is described by
\[
U_{\mathrm{PBS}} =
\begin{pmatrix}
1 & 0 & 0 & 0 \\
0 & 0 & 0 & i \\
0 & 0 & 1 & 0 \\
0 & i & 0 & 0
\end{pmatrix}.
\]

\medskip

A mirror imparts a relative phase $\phi$ between horizontal and vertical polarizations. In our setup, both mirrors affecting the two spatial modes introduce the same polarization-dependent phase on the vertical component. The corresponding unitary operation is expressed as
\[
U_{\mathrm{Mirror}} =
\begin{pmatrix}
1 & 0 & 0 & 0 \\
0 & e^{i\phi} & 0 & 0 \\
0 & 0 & 1 & 0 \\
0 & 0 & 0 & e^{i\phi}
\end{pmatrix}.
\]

The full setup shown in Fig.~2 of the main text was characterized using classical light. Laser light was injected into inputs 1 and 2 with four different polarizations: $\ket{\mathrm{Input}}$=$\ket{H}$,$\ket{V}$,$\ket{D}$ and $\ket{A}$, and the output state was reconstructed by performing single-qubit polarization quantum state tomography in both of the spatial outputs. By fitting the experimental data with a theoretical model incorporating all the key setup parameters, such as wave plate angles and phase shifts, we reconstructed the experimentally realized unitary transformation. The retrieved parameters generate a unitary evolution which, for ideal input photons, would generate a state with a fidelity of $0.988\pm0.004$ with respect to the three-mode NOON state in Eq.(1) of the main text.

\subsection{S1.2. Integrated optical circuit}
\label{supp:optical_circuit}
Our experimental implementation employs a bulk-optical circuit; however, the same transformation can be realized within an integrated photonic platform. For an input state of three indistinguishable single photons prepared in separate spatial modes $\ket{1,1,1,0}$, 
the following unitary also produces the desired state
\begin{align}\label{matrix}
{\color{blue}
U'=
\renewcommand{\arraystretch}{1.2}
\begin{pmatrix}
\frac{1}{\sqrt{2}} & \frac{1}{\sqrt{2}} & 0 & 0 \\
\frac{1}{\sqrt{6}} & -\frac{1}{\sqrt{6}} & -\sqrt{\frac{2}{3}} & 0 \\
-\frac{i}{\sqrt{6}} & \frac{i}{\sqrt{6}} & -\frac{i}{\sqrt{6}} & \frac{i}{\sqrt{2}} \\
\frac{1}{\sqrt{6}} & -\frac{1}{\sqrt{6}} & \frac{1}{\sqrt{6}} & \frac{1}{\sqrt{2}}
\end{pmatrix},
}
\end{align}
and it can be implemented by three beam splitters (BS) in sequence
\begin{equation}
    U'= U_{\mathrm{BS3}}U_{\mathrm{BS2}}U_{\mathrm{BS1}}.
\end{equation}
A beam splitter acting on two modes is described by
\begin{align} \label{BS}
U_{\text{BS}} =
\begin{pmatrix}
\cos(\theta) & \sin(\theta) \\
-\sin(\theta) & \cos(\theta) \\
\end{pmatrix}.
\end{align}
The first element of the integrated circuit is a balanced (50:50) beam splitter and acts on paths 1 and 2, corresponding to $\theta=\pi/4$:
\begin{align}
U_{\mathrm{BS1}} =
\begin{pmatrix}
\frac{1}{\sqrt{2}} & \frac{1}{\sqrt{2}} & 0 & 0 \\
-\frac{1}{\sqrt{2}} & \frac{1}{\sqrt{2}}  & 0 & 0 \\
0 & 0 & 1 & 0 \\
0 & 0 & 0 & 1
\end{pmatrix}.
\end{align}
The second element is an \textcolor{blue}{unbalanced} beam splitter with $\theta= \arccos(-\frac{1}{\sqrt{3}})$ acting on paths 2 and 3, with the matrix given by
\begin{align}
U_{\mathrm{BS2}} =
\begin{pmatrix}
1 & 0 & 0 & 0 \\
0 & -\frac{1}{\sqrt{3}} & -\sqrt{\frac{2}{3}} & 0 \\
0 & \sqrt{\frac{2}{3}} & -\frac{1}{\sqrt{3}} & 0 \\
0 & 0 & 0 & 1
\end{pmatrix}.
\end{align}
And, finally, $U_{\mathrm{BS3}}$ is another balanced beam splitter acting on paths 3 and 4
\begin{align}
U_{\mathrm{BS3}} =
\begin{pmatrix}
1 & 0 & 0 & 0 \\
0 & 1 & 0 & 0 \\
0 & 0 & \frac{1}{\sqrt{2}} & \frac{1}{\sqrt{2}} \\
0 & 0 & -\frac{1}{\sqrt{2}} & \frac{1}{\sqrt{2}}
\end{pmatrix}.
\end{align}

\section{S2. Measurement protocol}

\subsection{S2.1. Fidelity estimation}
We consider as target the three-mode, two-photon NOON state
\begin{equation}
\ket{\psi_3^2} = \frac{1}{\sqrt{3}} \left( \ket{2,0,0} + e^{i\alpha_1} \ket{0,2,0} + e^{i\alpha_2} \ket{0,0,2} \right),
\label{eq:supp:noon}
\end{equation}
where $\alpha_1, \alpha_2 \in [0, 2\pi)$ are real phase parameters.
Given an experimentally prepared state $\rho$, the fidelity with respect to the target is 
\begin{equation}
    F = \bra{\psi_3^2} \rho \ket{\psi_3^2}.
\end{equation}
A straightforward expansion gives
\begin{align}
F(\alpha_1, \alpha_2) = \frac{1}{3} \Big[& P_{200} + P_{020} + P_{002} + 2\,\mathrm{Re}\left( e^{-i\alpha_1}\rho_{200,020} + e^{-i\alpha_2}\rho_{200,002} + e^{-i(\alpha_2-\alpha_1)}\rho_{020,002} \right) \Big],
\label{eq:supp:fidelity_phases}
\end{align}
where $P_{n_{A} n_{B} n_{C}} \equiv \rho_{n_{A} n_{B} n_{C}, n_{A} n_{B} n_{C}}$ denote the population terms.
Using $\mathrm{Re}(e^{i\theta}z) = \mathrm{Re}(z)\cos\theta - \mathrm{Im}(z)\sin\theta$ for $z\!\in\!\mathbb{C}$, we obtain an explicit expression in terms of the real and imaginary parts of the coherence elements,
\begin{align}
F(\alpha_1, \alpha_2) = \frac{1}{3} \Big\{& P_{200} + P_{020} + P_{002}  + 2\big[\mathrm{Re}(\rho_{200,020})\cos\alpha_1 + \mathrm{Re}(\rho_{200,002})\cos\alpha_2  + \mathrm{Re}(\rho_{020,002})\cos(\alpha_2-\alpha_1) \big] \nonumber\\
& + 2\big[\mathrm{Im}(\rho_{200,020})\sin\alpha_1 + \mathrm{Im}(\rho_{200,002})\sin\alpha_2 + \mathrm{Im}(\rho_{020,002})\sin(\alpha_2-\alpha_1) \big] \Big\}.
\label{eq:supp:fidelity_full}
\end{align}

The three populations are experimentally accessible using the pseudo-photon-number-resolving detection method outlined in the main text. 
The off-diagonal elements $\rho_{200,020}$, $\rho_{200,002}$, and $\rho_{020,002}$ encode the pairwise two-mode coherences of the state and require dedicated interference measurements, which are discussed in the next subsection.

Since our target state is defined up to two relative phases $\alpha_1$ and $\alpha_2$, we adopt an optimization strategy to find the best fit over all possible phase values. 
For a given set of experimentally measured density matrix elements, the estimated fidelity is then given by 
\begin{equation}
F = \max_{\alpha_1, \alpha_2} F(\alpha_1, \alpha_2).
\label{eq:fidelity_optimized}
\end{equation} 
This optimization can be performed numerically using standard gradient-based or global optimization algorithms.  
The optimal phases $(\alpha_1^{\text{opt}}, \alpha_2^{\text{opt}})$ obtained from the best fidelity fit provide information about the relative phases present in the experimentally prepared state, and return the fidelity between the experimentally prepared state $\rho$ and the closest multi-mode NOON state.

\subsection{S2.2. Measurement strategy for coherence extraction}
\label{sec:supp:MeasurementStrategy}
\begin{figure*}
    \centering
    \includegraphics[width=0.65\textwidth]{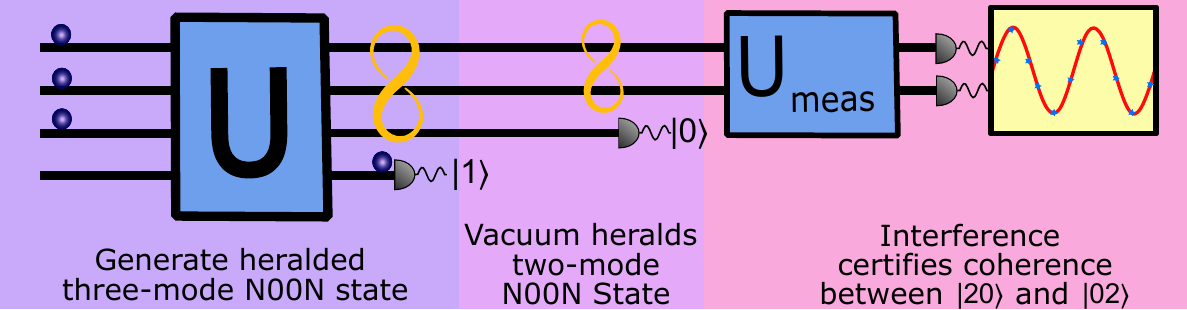}
    \caption{Schematic summary of the measurement sequence used to extract the interference fringes presented in Fig~3 of the main text.}
    \label{fig: CoherenceMeas}
\end{figure*}

The experimental characterization of the three-mode state $\rho$ relies on a sequence of projective measurements designed to extract the two-mode coherence terms of the density matrix. Fig.~\ref{fig: CoherenceMeas} provides a schematic overview of the protocol, which proceeds as follows: 
\begin{enumerate}
    \item \textbf{Vacuum projection:} One of the three modes is projected onto the vacuum state $\ket{0}$, effectively reducing the system to a two-mode subspace.
    \item \textbf{Unitary transformation:} The remaining two modes undergo a parameterized unitary transformation $U_{\mathrm{meas}}(\theta)$, implemented by a sequence of wave plates, before entering a polarizing beam splitter (PBS).
    \item \textbf{Coincidence detection:} The probability of the simultaneous detection of one photon in each output port of the PBS is recorded as a function of the rotation angle $\theta$.
\end{enumerate}
This procedure is repeated three times, projecting each of the three modes onto vacuum in turn, and yielding three coincidence curves, $C_{BC}(\theta)$, $C_{AC}(\theta)$, and $C_{AB}(\theta)$, corresponding to vacuum projection of modes $A$, $B$, and $C$, respectively.
Consider now projecting mode $k \in \{A, B, C\}$ onto the vacuum state.
The (unnormalized) state of the remaining modes is
\begin{equation}
\rho^{\,(ij)} = \bra{0}_{k}\, \rho \, \ket{0}_{k},
\end{equation}
with $\{i,j,k\}$ a permutation of $\{A,B,C\}$.  
The corresponding normalized state is
\begin{equation}
\rho'^{\,(ij)}
= \frac{\rho^{\,(ij)}}{\rho^{\,(ij)}_{20,20} + \rho^{\,(ij)}_{02,02} + \rho^{\,(ij)}_{11,11}},
\label{eq:supp:post_meas_state}
\end{equation}
Explicitly, for each choice of projected mode, we define:
\begin{subequations}
\begin{align}
\text{Mode } A \text{ projected:} \quad \rho^{(BC)}_{n_B n_C, m_B m_C} &= \rho_{0\, n_B n_C, 0\, m_B m_C}, \\
\text{Mode } B \text{ projected:} \quad \rho^{(AC)}_{n_A n_C, m_A m_C} &= \rho_{n_A 0\, n_C, m_A 0\, m_C}, \\
\text{Mode } C \text{ projected:} \quad \rho^{(AB)}_{n_A n_B, m_A m_B} &= \rho_{n_A n_B 0, m_A m_B 0}.
\end{align}
\label{eq:supp:density_correspondence}
\end{subequations}
 
The unitary transformation $U(\theta)$ acting on modes $i$ and $j$ is implemented through the wave plate sequence 
\begin{equation}
U(\theta) = \text{HWP}(\theta)\, \text{QWP}(\pi/4),
\label{eq:supp:wave_plate_config}
\end{equation} 
where $\text{QWP}(\phi)$ and $\text{HWP}(\phi)$ denote quarter-wave and half-wave plates, respectively, with fast axes oriented at angle $\phi$ relative to a laboratory-fixed reference frame. 
This configuration mixes the real and imaginary parts of the coherence term $\rho^{(ij)}_{20,02}$, enabling their simultaneous extraction from the measured fringe. 
After the combined unitary transformation and PBS operation, the density matrix in the measurement basis is 
\begin{equation}
\tilde{\rho}^{(ij)}(\theta) = U_{\text{PBS}} \, U(\theta) \, \rho'^{\,(ij)} \, U(\theta)^\dagger \, U_{\text{PBS}}^\dagger,
\label{eq:supp:transformed_density}
\end{equation} 
and the coincidence probability of detecting one photon in each output port is 
\begin{equation}
C_{ij}(\theta)  
= \bra{1,1} \tilde{\rho}^{(ij)}(\theta) \ket{1,1}.
\label{eq:supp:coincidence_prob}
\end{equation} 
 
For the wave plate configuration in Eq.~\eqref{eq:supp:wave_plate_config}, $C_{ij}(\theta)$ has the sinusoidal form 
\begin{equation}
C_{ij}(\theta) = \frac{A_{ij}}{2} + \frac{V_{ij}}{2} \cos(8\theta + \varphi_{ij}),
\label{eq:supp:coincidence_form}
\end{equation} 
where $A_{ij}$ is the offset, $V_{ij}$ the visibility, and $\varphi_{ij}$ the fringe phase. 
These parameters are related to density matrix elements via 
\begin{subequations}\label{eq:supp:coefficients}
\begin{align}
A_{ij} &= \frac{\rho^{(ij)}_{20,20} + \rho^{(ij)}_{02,02}}{\rho^{(ij)}_{20,20} + \rho^{(ij)}_{02,02} + \rho^{(ij)}_{11,11}}, \label{eq:supp:offset} \\
V_{ij} &= \frac{ 2 |\rho^{(ij)}_{20,02}|}{\rho^{(ij)}_{20,20} + \rho^{(ij)}_{02,02} + \rho^{(ij)}_{11,11}}, \label{eq:supp:visibility}\\
\varphi_{ij} &= \mathrm{arg}(\rho^{(ij)}_{20,02});
\end{align}
\end{subequations} 
with $\rho^{\,(ij)}_{20,02} = |\rho^{\, (ij)}_{20,02}|\, e^{-i \varphi_{ij}}$.  
The visibility $V_{ij}$ then encodes the magnitude of the two-photon coherence, while the phase $\varphi_{ij}$ relates to its complex argument. 
Since $\rho'^{\,(ij)}$ is normalized, $\rho'^{\,(ij)}_{20,20} + \rho'^{\,(ij)}_{02,02} + \rho'^{\,(ij)}_{11,11} = 1$, we may equivalently write $A_{ij} = 1 - \rho'^{\,(ij)}_{11,11}$, showing that the offset $A_{ij}$ reflects the population of the $\ket{1,1}$ component. 
 
Thus, by fitting the measured coincidence curves $C_{ij}(\theta)$ to Eq.~\eqref{eq:supp:coincidence_form}, we can extract both the magnitude and phase of the coherence element $\rho^{(ij)}_{20,02}$ for each bipartition, providing the information required for the fidelity analysis in the main text.

\subsection{S2.3. Fidelity bounds from coherence measurements without populations}
\label{supp:coherence}

The measurement protocol described in the previous subsection provides direct access to the two-mode coherence elements of the three-mode state $\rho$. 
Although our experiment additionally provides pseudo-photon-number–resolved population measurements, it is instructive to also consider a more conservative scenario in which population information is unavailable or unreliable.   
We show here that the coincidence measurements alone already determine \emph{rigorous upper and lower bounds} on the fidelity with respect to the target three-mode NOON state in Eq.~\eqref{eq:supp:noon}. 
 
The key observation is that the definitions in Eqs.~\eqref{eq:supp:offset} and \eqref{eq:supp:visibility} when combined with the positivity constraint  
\begin{align}\label{eq:supp:cohineq}
    |\rho_{ij}|^2 \leq \rho_{ii}\rho_{jj},
\end{align} 
valid for any density matrix, impose nontrivial restrictions on the possible population distributions consistent with the observed data. 
To formalize this approach, we introduce a convenient notation for the relevant three-mode populations,
\begin{subequations}
\begin{align}
P_A &= \rho_{200,200}, \quad P_B = \rho_{020,020}, \quad P_C = \rho_{002,002}, \\
P_{AB} &= \rho_{110,110}, \quad P_{AC} = \rho_{101,101}, \quad P_{BC} = \rho_{011,011};
\end{align}
\end{subequations}
and define dimensionless population ratios,
\begin{equation}
r_B = \frac{P_B}{P_A}, \quad r_C = \frac{P_C}{P_A}, \quad 
r_{AB} = \frac{P_{AB}}{P_A}, \quad r_{AC} = \frac{P_{AC}}{P_A}, \quad r_{BC} = \frac{P_{BC}}{P_A}.
\end{equation}
Using these ratios, the offsets and visibilities obtained from the three coincidence curves may be written as
\begin{subequations}
\begin{align}
A_{BC} &= \frac{r_B + r_C}{r_B + r_C + r_{BC}}, &
V_{BC} &= \frac{2|\rho_{020,002}|}{P_B + P_C + P_{BC}}, \\
A_{AB} &= \frac{1 + r_B}{1 + r_B + r_{AB}}, &
V_{AB} &= \frac{2|\rho_{200,020}|}{P_A + P_B + P_{AB}}, \\
A_{AC} &= \frac{1 + r_C}{1 + r_C + r_{AC}}, &
V_{AC} &= \frac{2|\rho_{200,002}|}{P_A + P_C + P_{AC}}.
\end{align}
\end{subequations}
Applying the inequality in Eq.~\eqref{eq:supp:cohineq} to each relevant coherence term yields three constraints on the population ratios,
\begin{subequations}
\label{eq:supp:constraints}
\begin{align}
\frac{V_{AB}^{2}}{A_{AB}^{2}}(1 + r_{B})^{2} &\le 4r_{B}, \\
\frac{V_{AC}^{2}}{A_{AC}^{2}}(1 + r_{C})^{2} &\le 4r_{C}, \\
\frac{V_{BC}^{2}}{A_{BC}^{2}}(r_{B}+r_{C})^{2} &\le 4r_{B}r_{C}.
\end{align}
\end{subequations}
These inequalities define the feasible region $\mathcal{R}$ in $(r_B,r_C)$ space compatible with the observed visibilities and offsets, independently of any direct population measurement.
Furthermore, the fidelity in Eq.~\eqref{eq:supp:fidelity_full} can be expressed as
\begin{align}\label{eq:supp:fidelity_observables}
F(\alpha_1, \alpha_2) = \frac{P_A}{3} \Big[1 + r_B + r_C 
&+ \frac{V_{AB}}{A_{AB}}(1 + r_B)\cos(\alpha_1 - \varphi_{AB})
+ \frac{V_{AC}}{A_{AC}}(1 + r_C )\cos(\alpha_2 - \varphi_{AC})\nonumber\\
& + \frac{V_{BC}}{A_{BC}}(r_B + r_C)\cos(\alpha_2 - \alpha_1 - \varphi_{BC})\Big],
\end{align}
where $\alpha_1$ and $\alpha_2$ are the two relative phases of the state in Eq.~\eqref{eq:supp:noon}, and $P_A$ can be written as
\begin{align}\label{eq:supp:pa}
    P_A = \left[ 1 + (1 + r_B)\left( \frac{1}{A_{AB}} -1 \right)  + (1 + r_C)\left( \frac{1}{A_{AC}} -1 \right) +  \frac{r_B + r_C}{A_{BC}}\right]^{-1}.
\end{align}
As above, the phases must be fitted to obtain the maximum achievable fidelity consistent with the data.
To simplify this optimization, we combine the first two cosine terms using standard trigonometric identities.
Defining
\begin{equation}
S_A = \frac{V_{AB}}{A_{AB}}(1+r_B),\qquad
S_C = \frac{V_{AC}}{A_{AC}}(1+r_C),
\end{equation}
and letting
\begin{equation}
\Delta = \alpha_{2} - \alpha_{1} + \varphi_{AB} - \varphi_{AC},
\end{equation}
we obtain
\begin{align}
S_A\cos(\alpha_{1}-\varphi_{AB})
+ S_C\cos(\alpha_{2}-\varphi_{AC})
= R\cos(\alpha_{1}-\varphi_{AB}+\phi),
\end{align}
where
\begin{equation}
R = \sqrt{S_{A}^{2}+S_{C}^{2}+2S_{A}S_{C}\cos\Delta},
\qquad
\phi = \arctan\!\left[
\frac{S_{C}\sin\Delta}{S_{A}+S_{C}\cos\Delta}
\right].
\end{equation}
Maximizing $F$ over $\alpha_{1}$ is now straightforward by choosing $\alpha_{1} = \varphi_{AB}-\phi$, which yields
\begin{equation}
\bar{F}(\delta)
= \frac{P_A}{3}
\left[
1 + r_{B} + r_{C} 
+ R
+ \frac{V_{BC}}{A_{BC}}(r_{B}+r_{C})
  \cos(\delta-\varphi_{BC})
\right],
\end{equation}
where $\delta = \alpha_{2}-\alpha_{1}$ is the remaining free phase, and $P_A$ is given by Eq.~\eqref{eq:supp:pa}. 
For fixed $(r_B,r_C)$, the maximum fidelity is obtained by optimizing $\bar{F}(\delta)$ over~$\delta$.
The full optimization used to compute fidelity bounds is therefore:
\begin{enumerate}
\item Extract $V_{ij}$, $\varphi_{ij}$ and $A_{ij}$ from the measured
coincidence fringes.
\item Determine the feasible region $\mathcal{R}$ in $(r_B,r_C)$ satisfying the
constraints in Eqs.~\eqref{eq:supp:constraints}.
\item For each $(r_B,r_C)\in\mathcal{R}$, compute
\[
F_{\max}(r_B,r_C) = \max_{\delta}\,\bar{F}(\delta).
\]
\item The fidelity bounds follow as
\begin{equation}
F_{\mathrm{lower}}
= \min_{(r_B,r_C)\in\mathcal{R}} F_{\max}(r_B,r_C),\qquad
F_{\mathrm{upper}}
= \max_{(r_B,r_C)\in\mathcal{R}} F_{\max}(r_B,r_C).
\end{equation}
\end{enumerate}
This procedure provides rigorous fidelity bounds $F \in [F_{\mathrm{lower}}, F_{\mathrm{upper}}]$ using \emph{only} the information contained in the coincidence measurements.
When population measurements are available, the population ratios $(r_B,r_C)$ are fixed uniquely, collapsing the bounds to a single fidelity value.
Applying this method to the measured fringes shown in
Fig.~3 of the main text yields $ F \in [0.818,\,0.836]$, with experimental uncertainties propagated throughout the optimization.

\subsection{S2.4. Genuine tripartite entanglement threshold for the three-mode two-photon NOON state}

To certify genuine tripartite entanglement, we determine the maximal fidelity between a target pure tripartite state $|\psi\rangle$ and any biseparable state.
To obtain this general bound, we follow the procedure described in the Appendix of  Ref.~\cite{SMBourennane2004}.
For completeness, and to keep this work self-contained, we reproduce the argument here.
A mixed state $\sigma$ on parties $A,B,C$ is biseparable if it can be written as a convex combination of pure states that are separable with respect to at least one bipartition ($A|BC$, $B|AC$, or $C|AB$)~\cite{SMBourennane2004}.  
However, since any mixed state is a convex combination of pure states and the fidelity $F(|\psi\rangle,\sigma)=\langle\psi|\sigma|\psi\rangle$ is linear, the maximum fidelity with respect to biseparable states is always achieved by a pure biseparable state.
We therefore define the biseparable fidelity bound as
\begin{equation}\label{eq:supp:fbs_def}
    F_{\mathrm{bs}}
    := \max_{|\phi\rangle\in\mathcal{B}}
       |\langle\phi|\psi\rangle|^{2},
\end{equation}
where $\mathcal{B}$ denotes the set of pure biseparable states.
Thus, any state $\rho$ satisfying $F(|\psi\rangle,\rho) > F_{\mathrm{bs}}$ is certified to be genuinely tripartite entangled.
To determine $F_{\mathrm{bs}}$, we first fix a bipartition, say $A|BC$, and choose orthonormal bases $\{|i\rangle_{A}\}$ and $\{|j\rangle_{BC}\}$ for the corresponding subsystems.  
In this basis, any normalized pure tripartite state $|\psi\rangle$ can be written as
\begin{equation}\label{eq:supp:psi_coeff}
    |\psi\rangle = \sum_{i,j} C_{ij}\, |i\rangle_{A}\,|j\rangle_{BC},
\end{equation}
for some complex coefficient matrix $C$.
Any pure product state across the same bipartition has the form
\begin{equation}
    |\phi\rangle = |a\rangle |b\rangle
    = \sum_{i,j} a_{i} b_{j}\, |i\rangle_{A}\,|j\rangle_{BC},
\end{equation}
where the vectors $\mathbf{a} = (a_1, a_2, \ldots)^{T}$ and $\mathbf{b} = (b_1, b_2, \ldots)^{T}$ are normalized, $\sum_{i}|a_{i}|^{2} = \sum_{j}|b_{j}|^{2} = 1$.
The overlap between the product state $|\phi\rangle$ and the target state $|\psi\rangle$ is then
\begin{equation}
    \langle\phi|\psi\rangle
    = \sum_{i,j} a_{i}^{*} C_{ij} b_{j}^{*}
    = \mathbf{a}^{\dagger} C\, \mathbf{b}^{*}.
    \label{eq:overlap}
\end{equation}
Maximizing the fidelity $|\langle\phi|\psi\rangle|^{2}$ over product states $|\phi\rangle$ is therefore equivalent to maximizing
$|\mathbf{a}^{\dagger} C\, \mathbf{b}^*|^{2}$ over all normalized vectors $\mathbf{a}$ and $\mathbf{b}$.
To obtain a useful upper bound, we perform a singular value decomposition $C = U \Sigma V^{\dagger}$, where $U$ and $V$ are unitary and $\Sigma = \mathrm{diag}(\sigma_{1},\sigma_{2},\ldots)$ collects the singular values $\sigma_{k} \ge 0$ of $C$.
Introducing the normalized vectors $\mathbf{x} = U^{\dagger} \mathbf{a}$ and $\mathbf{y} = V^{\dagger} \mathbf{b}^*$,
we obtain
\begin{align}
    |\mathbf{a}^{\dagger} C\, \mathbf{b}^*|^{2}
    &= |\mathbf{a}^{\dagger}\, U \Sigma V^{\dagger}\, \mathbf{b}^*|^{2}
     = |\mathbf{x}^{\dagger} \Sigma\, \mathbf{y}|^{2}  \\
    &= \biggl|\sum_{k} \sigma_{k}\, x_{k}^{*} y_{k}\biggr|^{2}
     \;\le\; \sigma_{\max}^{2}\,
        \biggl|\sum_{k} x_{k}^{*} y_{k}\biggr|^{2},
\end{align}
where $\sigma_{\max} = \max_{k}\sigma_{k}$ is the largest singular value of $C$.
Further, by the Cauchy–Schwarz inequality,
\begin{equation}
    \biggl|\sum_{k} x_{k}^{*}\, y_{k}\biggr|^2 \le \sum_{k} |x_{k}|^2\,  \sum_{k} |y_{k}|^2  = 1,
\end{equation}
and the fidelity then satisfies
\begin{equation}
    |\langle\phi|\psi\rangle|^{2}  = |\mathbf{a}^{\dagger} C\, \mathbf{b}^*|^{2}
    \le \sigma_{\max}^{2}.
    % \label{eq:max_fid_bipartition}
\end{equation}
Therefore, for the bipartition $A|BC$,
\begin{equation}
    \max_{|\phi\rangle\in A|BC}
       |\langle\phi|\psi\rangle|^{2} = \max_{k} \sigma_{k}^{2}.
    \label{eq:max_fid_bipartition}
\end{equation}
The same optimization can be carried out for the remaining bipartitions $B|AC$ and $C|AB$.  We therefore obtain the biseparable fidelity bound as
\begin{equation}
    F_{\mathrm{bs}}
    = \max\!\left\{
        \max_{k}\sigma^{2}_{k}(A|BC),\;
        \max_{k}\sigma^{2}_{k}(B|AC),\;
        \max_{k}\sigma^{2}_{k}(C|AB)
      \right\},
\end{equation}
that is, the maximum singular value squared across all three possible bipartitions of the system.
Finally, we recall that the singular values of the coefficient matrix $C$ appearing in Eq.~\eqref{eq:supp:psi_coeff} are precisely the Schmidt coefficients of $|\psi\rangle$ with respect to the chosen bipartition.
Then, $F_{\mathrm{bs}}$ is simply the largest squared Schmidt coefficient of $|\psi\rangle$ when considering all bipartitions of the tripartite Hilbert space.
A fidelity exceeding this value therefore certifies genuine tripartite entanglement.

We now apply this general result to the three-mode two-photon NOON state $|\psi_3^2\rangle$ in Eq.~\eqref{eq:supp:noon}, which is symmetric under permutations of the modes.  
For the partition $A|BC$ we introduce the orthonormal states
\begin{equation}
|\varphi_{1}\rangle
=\frac{1}{\sqrt{2}} \left(e^{i\alpha_{1}}|20\rangle+e^{i\alpha_{2}}|02\rangle\right),
\qquad
|\varphi_{2}\rangle = |00\rangle,
\end{equation}
so that
\begin{equation}
|\psi_3^2\rangle
=\sqrt{\frac{2}{3}}\;|0\rangle\,|\varphi_{1}\rangle
+\sqrt{\frac{1}{3}}\;|2\rangle\,|\varphi_{2}\rangle.
\end{equation}
Then, the coefficient matrix $C$ will be given by
\begin{align}
C = 
\begin{pmatrix}
    \sqrt{\frac{2}{3}} & 0 \\
    0 & \sqrt{\frac{1}{3}}
\end{pmatrix},
\end{align}
and the Schmidt coefficients for $A|BC$ are therefore $\sqrt{2/3}$ and $\sqrt{1/3}$.  
By symmetry, the same Schmidt spectrum holds for the bipartitions $B|AC$ and $C|AB$.
Thus, the largest squared Schmidt coefficient over all bipartitions is
\begin{equation}
    F_{\mathrm{bs}} = \frac{2}{3},
\end{equation}
independent of the phases $\alpha_{1}$ and $\alpha_{2}$. 
Any state $\rho$ with $F(|\phi\rangle,\rho)>2/3$ is therefore genuinely tripartite entangled.

\section{S3. Experimental details}
\subsection{S3.1. Input state source design}\label{subsec:source}
The input state $\ket{1,1,1,0}$ is generated using a multi-photon source based on the design of Ref.~\cite{Loophole-free2015, EPRsteering2018}, which produces four single photons in four modes from two simultaneous type-II spontaneous parametric down-conversion (SPDC) events within a beam-displacer interferometer. A horizontally polarized picosecond 775 nm pump beam first passes through a half-wave plate (HWP) that transforms its polarization to an equal superposition of H and 
V components, which are then spatially separated into two horizontally displaced beams by the first beam displacer (BD). Two subsequent HWPs rotate both beams to horizontal polarization, suitable for type-II down-conversion, while matching their optical paths. These beams pump a 15-mm-long periodically poled potassium titanyl phosphate (ppKTP) crystal in two distinct locations, allowing for degenerate type-II SPDC. A second BD vertically separates the signal and idler photons, producing four down-converted photons in four distinct beams. Three additional HWPs then rotate the polarizations so that single photons in the two left beams are H-polarized and the two right beams are V-polarized. This configuration allows the signal photons from the two down-conversion regions to be recombined into one path by the third BD, and likewise for the idler photons. At the output of the BD interferometer, two vertically separated beams remain, each containing a pair of orthogonally polarized photons. The generated state can be written as $\ket{1_{\mathrm{H}},1_{\mathrm{V}}}_{\mathrm{top}}\ket{1_{\mathrm{H}},1_{\mathrm{V}}}_{\mathrm{bottom}}$. A D-shaped mirror separates the propagation paths of the top and bottom beams, which are subsequently collimated, spectrally filtered with a 1-nm band-pass filter centered at 1550 nm, and coupled into single-mode fibers. The V-polarized photon in the bottom beam is detected with an SNSPD, while the remaining H-polarized photon, is sent to input 2 of the state generation setup. The bottom-beam pair, $\ket{1_{\mathrm{H}},1_{\mathrm{V}}}_{\mathrm{bottom}}$, is sent to input 1. This procedure yields the separable input state used in our experiment: $\ket{1_{\mathrm{H}},0_{\mathrm{V}}}_{\mathrm{Input2}}\ket{1_{\mathrm{H}},1_{\mathrm{V}}}_{\mathrm{Input1}}$.

\begin{figure}[t]
    \centering
    \includegraphics[width=0.7\linewidth]{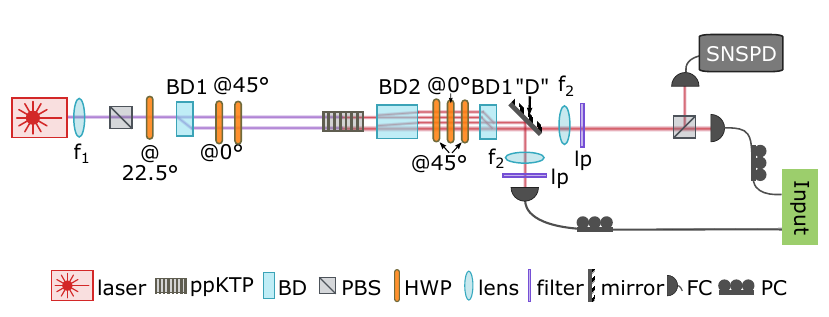}
    \caption{\textbf{Input state generation}. Schematic of the source, as detailed in Sec.~\ref{subsec:source}. Abbreviations: ppKTP – periodically poled potassium titanyl phosphate; BD – (polarizing) beam displacer; PBS – polarizing beam splitter; HWP – half-wave plate; QWP – quarter-wave plate; FC – (single-mode) fiber coupler; BS – (fiber) beam splitter; PC – (fiber) polarization controller; SNSPD – superconducting nanowire single-photon detector; lp – long-pass filter; bp – band-pass filter; “D” – D-shaped mirror with a horizontal cut (not visible in top view). BD2 provides a vertical beam displacement, illustrated in the schematic by the slight vertical separation of beam pairs (source design based on \cite{EPRsteering2018}).}
    \label{fig:placeholder}
\end{figure}

\subsection{S3.2. Measurements of interference fringes and populations}

\begin{figure}
    \centering
    \includegraphics[width=0.5\linewidth]{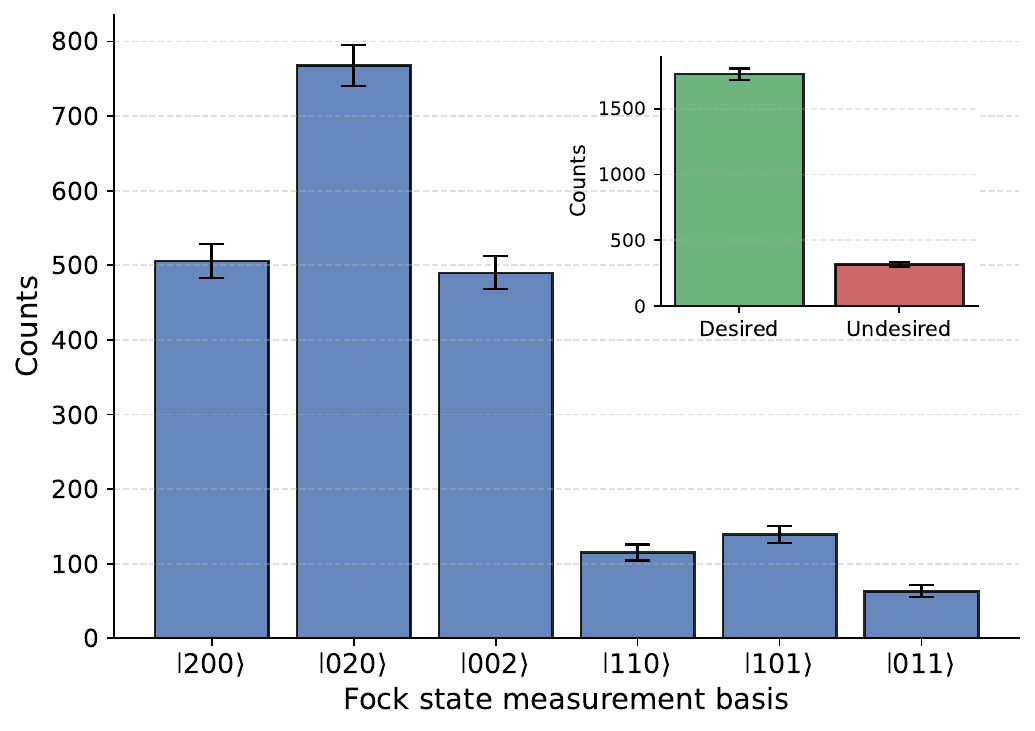}
    \caption{\textbf{Populations}: Photon-number populations measured in the Fock basis. Dominant contributions arise from the NOON components: 
     $\ket{200}$, $\ket{020}$ and $\ket{002}$. Their sum and the sum of the undesired events ($\ket{110}$, $\ket{101}$,  and $\ket{011}$) 
     are shown for comparison in the inset.} 
    \label{fig: populations}
\end{figure}

The interference fringes (Fig.~3 of the main text) were measured using the pseudo-photon-number-resolution detection mentioned in the main text. The four-fold coincidence probabilities were normalized by taking into account the $\ket{1,1}$, $\ket{2,0}$ and $\ket{0,2}$ terms:
\begin{equation}
    P_{\ket{1,1}} = \frac{\text{Counts}_{\ket{1,1}}}{\text{Counts}_{\ket{1,1}} + \text{Counts}_{\ket{2,0}} + \text{Counts}_{\ket{0,2}} }\: .
\end{equation}
The experimentally observed visibilities for the three sets of probability curves are $(81.9\pm3.6)$\% for modes (A,B), $(92.0\pm2.8)$\% for modes (B,C) and $(81.2\pm4.0)$\% for modes (A,C) of the state $\ket{\psi_3^2}$. The centers of oscillation for the three curves in Fig.~3 of the main text, obtained by fitting the data with Eq.~\eqref{eq:supp:coincidence_form}, are $0.467\pm0.005$ ($C_{AB}$), $0.478\pm0.007$ ($C_{BC}$), and $0.480\pm0.006$ ($C_{AC}$). These values are close to the ideal of 0.5 and the slight reduction is due to the presence of undesired terms in the generated state. The sum of the measured populations is $0.847\pm0.038$ and the recorded counts are shown in Fig.~\ref{fig: populations}.
\subsubsection{1. Experimental imperfections}
To better understand the different contributions affecting the experimental fidelity, we delineate between systematic imperfections that reduce its mean value and statistical fluctuations that determine its uncertainty. 
In our experiment, the primary systematic effect arises from imperfect quantum interference between the three input photons. 
% The reduction in fidelity is mainly due to the residual distinguishability among the three photons.  
Even with a perfect experimental setup implementing the ideal unitary transformation, three photons with an indistinguishability of 0.85 (our measured independent source visibility) would yield a heralded state with a fidelity of $0.869\pm0.007$, using an orthogonal bad-bit model~\cite{jeff}. This contribution may in fact be slightly overestimated in our conservative model, which assigns a fixed distinguishability between all photons. This is because in the model we use the Hong–Ou–Mandel visibility we obtained experimentally for photons from different sources, whereas two of the three photons originate from the same source and are therefore expected to exhibit a higher HOM visibility. Apart from this major contribution, we expect additional contributions from the imperfect unitary, slow source fluctuations between measurements (primarily due to laser frequency and mode-locking instability), higher-order SPDC events, detector efficiency imbalance, etc.
% Systematic effects include: (i) imperfect quantum interference arising from partial distinguishability and spectral correlations leading to mixture of single photons, (ii) deviations of the implemented unitary transformation from the ideal one, (iii) slow drifts in the source affecting the photon generation rate, 
% and, (iv) the difference in the efficiencies of the detectors. 
These effects can lead to a reduction in achievable fidelity, even in the limit of infinite counting statistics. 
In contrast, the statistical uncertainty on the fidelity is due to the Poissonian uncertainties associated with photon-counting events and the source fluctuations during individual measurements (see Sec.~S5 of the SM).

\subsubsection{2. Technical specifications}

Our detectors had an average detection efficiency of $\sim 87.6\%$ and dark count rates on the order of 200 Hz. 
% However, the probability of measuring four-fold detections owing to the dark counts is negligible. %\sout{\textcolor{red}{We need to add the double pair emission from the SPDC, these create coincidences too!}}. 
% Triple-pair emissions from the SPDC source are estimated to contribute only 7.9\% to the undesired terms. 
The total detection system jitter was below 1 ns, so a 2.5 ns coincidence window was chosen, resulting in negligible dark counts. 
We estimate the noise from triple-pair emission and find that it is expected to contribute 7.9\% of the measured undesired terms ($\ket{1,1,0}$, $\ket{1,0,1}$ and $\ket{0,1,1}$). Therefore, triple-pair emission is not the primary reason for the fidelity reduction. 
The Sagnac interferometers were sufficiently stable such that no active stabilization was used during the experiment. Over a period of 23~hours, the measured phase drift was approximately $1.14$~deg, corresponding to only $\sim 0.45$~deg during the acquisition time of a single coincidence curve.
% The minimum of an experimentally measured curve shifted less than 5 degrees over a measurement period of 14 hours, corresponding to a phase drift smaller than $10^{-4}$ deg/s. 
% As discussed above, photon distinguishability is a primary contributor to the reduced state fidelity. 
% Another contribution arises from the experimental imperfection in implementing the ideal unitary operator. 
% We characterize our unitary with classical light and reconstruct a transformation that would generate a state with fidelity of $\approx 0.99$, if a perfect input state were available. However, this does not capture all imperfections of the apparatus.  
Although classical characterization reveals a near-ideal transformation (see Sec~S1.1 of the SM), simultaneous optimization of the spatial overlap between all modes was not possible. This was revealed by dependent HOM measurements, which showed visibility of $(91.6\pm0.4)\%$ in output 1 and $(98.2\pm0.2)\%$ in output 2 using the same input (inputs and outputs are illustrated in Fig.~1(b) of the main text).

\subsection{S3.3. Heralded state generation rate}
\subsubsection{1. Heralded state generation rate for our source}
We estimate the overall rate of heralded three-mode NOON state generation to be $8.1 \times 10^{-6}$ and $1.1 \times 10^{-5}$ per pump pulse with and without final photon counting,  respectively. The difference between these two estimates amounts to the detection of the target-state photons for the case of final photon counting. These estimates were inferred from experimental data as follows: We first characterized the source, starting from the measured single-photon counts of the ``trigger photon''. The trigger photon is the fourth photon, which does not participate in the state-generation circuit but whose detection is required to indicate the existence of its partner photon, one of the three input photons into the circuit. %the preparation of the correct input statea detector placed outside the unitary transformation stage. 
%This detector acts as a trigger monitoring the fourth photon, which does not participate in the state-generation circuit but whose detection is required to certify the preparation of the correct input state. 
We similarly recorded singles for the heralding detector, which registers the heralding photon within the three-photon state generation circuit. Comparing these two singles counts and taking into account the expected action of the unitary, we inferred the transmission efficiency of the state generation unitary transformation ($\eta_{\mathrm{eff}}=0.72$), under the assumptions that detection efficiencies are equal and that the two photon-pair sources generate the same number of pairs per pump pulse (as per design).
%From the measured singles of the external trigger detector and coincidence rates between the trigger detector and the heralding detector, and taking into account the expected action of the unitary transformation on one of the partner photons
Taking into account the expected action of the unitary transformation on one of the photons, and using the measured coincidences $C_{\mathrm{TH}}$ between the trigger detector and the heralding detector together with the singles $S_{\mathrm{T}}$ recorded by the trigger detector, we estimate an overall heralding (Klyshko) efficiency of $\eta_{\mathrm{\mathrm{H}}} \approx 0.14$. 
This value incorporates coupling losses, detector inefficiencies, and losses within the optical circuit used to implement our unitary.
Using $S_T$, which is not affected by circuit losses, we estimate the number of pairs generated to be $S_T\,\eta_{\mathrm{eff}}/\eta_{\textrm{H}}$. Dividing this value by the acquisition time ($1800\,\mathrm{s}$) and then by the pump laser repetition rate ($R_{\mathrm{pump}} = 80~\mathrm{MHz}$), we estimate the probability that the source generates a photon pair per pump pulse to be $p_{2} = 8.21 \times 10^{-3}$. 
%From here, we estimate the probability of generating a photon pair per pump pulse to be $p_{2} = 8.21 \times 10^{-3}$. 
The probability of producing the desired four photons (one pair from each of the two sources) per pump pulse is then $p_{4} = p_{2}^{2} = 6.7 \times 10^{-5}$.

Using these quantities, the expected four-fold coincidence rate associated with successful NOON-state generation is
\setcounter{equation}{50}
\begin{equation}
C_{\mathrm{4\text{-}folds}}^{\mathrm{NOON}} 
    = R_{\mathrm{pump}} \, p_{4} \, p_{\mathrm{U}} \,
    \, \eta_{\mathrm{BS}}^{2} \, \eta_{\mathrm{ext}} \, \eta_{\mathrm{H}}^{3}.
\end{equation}
Here, our four-fold coincidences have been normalized to mitigate the probabilistic nature of the pseudo-photon-number-resolving detection technique. The factor 
$p_{\mathrm{U}} = 1/4$ is the intrinsic success probability of the state generation scheme%$p_{\mathrm{dist}} = 1/3$ accounts for the equal probability of distributing the photons among the three output modes
. 
Furthermore, $\eta_{\mathrm{BS}} \approx 0.96$ denotes the manufacturer-specified transmission efficiency of the fiber beam splitters used to implement pseudo-photon-number resolution, and through which two photons pass. These fiber beam splitters are placed in each of the outputs A, B, and C of the unitary in Fig.~1(a) of the main text. 
The quantity $\eta_{\mathrm{H}} \approx 0.14$ accounts for the detection, coupling, and unitary circuit efficiency for each of the three photons that pass through the unitary circuit, while $\eta_{\mathrm{ext}} \approx 0.19$ accounts for the detection and coupling efficiencies of the fourth photon monitored outside the unitary. 
%The latter is higher due to the absence of unitary losses, consistent with $\eta_{\mathrm{ext}} \approx \eta_{\mathrm{H}} / 0.73$.
Substituting these values yields an expected four-fold coincidence detection rate for the heralded state generation of %$0.2146~\mathrm{Hz}$ per output mode, corresponding to a total rate of 
$0.644~\mathrm{Hz}$ or $8.1 \times 10^{-6}$ per pump pulse. 
This estimate roughly agrees with the experimentally observed heralded state-generation rate of $0.88\pm0.02~\mathrm{Hz}$ or $1.09 \times 10^{-5}$ per pump pulse.

Since the photon counting would usually be delayed until after the state is used, it also makes sense to estimate the state generation rate without final photon counting, i.e., prior to the detection of the two photons in the target state. For these purposes, the $C_{\mathrm{4\text{-}folds}}^{\mathrm{NOON}}$ rate is divided by the efficiencies associated with the detection of the two target photons (their detection efficiencies and the efficiencies of the fiber beam splitters used to implement pseudo-photon-number resolution). Doing this, and using a rough estimate of our experimental detector efficiency of $\eta_{\mathrm{detector}}\approx0.875$, we obtain the rate $C_{\mathrm{4\text{-}folds}}^{\mathrm{NOON}}/(\eta_{\mathrm{detector}}^2\eta_{\mathrm{BS}}^2) = 0.91~\mathrm{Hz}$, corresponding to $ 1.1 \times 10^{-5}$ per pump pulse.

\subsubsection{2. Heralded state generation rate for alternate photon sources}
As mentioned in the \textit{Discussion} section of the main text, three SPDC sources can be employed to eliminate the need for final photon counting. %given a low-loss optical circuit for the unitary. 
In our current configuration, six-photon generation would yield the target state at a rate of $\sim 2.4 \times 10^{-4}~\mathrm{Hz}$, where successful preparation of the input state is verified by detecting three photons externally (efficiency of 0.19) while their partner photons are routed into the unitary circuit. Of these three photons, the heralding photon, which is transmitted through the unitary and detected, has an efficiency of 0.14, while the target photons, which are only transmitted through the unitary but not detected, have an efficiency of 0.17. %and transmission efficiency of $\eta_\mathrm{eff}=0.72$ for each of the four modes. Heralding efficiency represents the efficiency for a single photon to propagate through the source and the optical circuit to the detectors, while transmission efficiency only accounts for the propagation through the optical circuit. The two photons of the three-mode NOON state do not need to be detected, and hence we do not need to account for the detection and fiber beam splitter efficiencies in those two modes. 
This rate could be drastically improved by reducing losses. For example, employing the same source in combination with a low-loss optical circuit (where the trigger photons have an efficiency of 0.7, the heralding photon 0.5, and the target photons 0.6), the expected generation rate for the triple-pair approach would increase to approximately $0.66~\mathrm{Hz}$. \\
%for example, using the same source with a low-loss unitary achieving $\eta_\mathrm{H} = 0.5$ and $\eta_\mathrm{eff}=0.9$ would increase the expected rate to $ \approx 0.3~\mathrm{Hz}$. \\ %provide the three efficiencies: 0.5, 0.55, 
%the rate of the target state can be increased drastically by increasing these efficiencies
% for example, using the same source with a low-loss optical circuit where the external photons are detected with an efficiency of 0.55, the heralding photon is detected with an efficiency of 0.5, and the photons in the target mode are transmitted through the optical circuit with an efficiency of 0.59 
\\
Alternatively, the overall preparation rate of the three-mode NOON state could be significantly increased by employing a state-of-the-art quantum dot source such as that reported in Ref.~\cite{DingQD2025}. In that work, Ding \emph{et al.} demonstrated a three-photon coincidence rate of approximately $4.5\,\mathrm{MHz}$, which would translate into a heralded state generation rate on the order of $\sim 3.1\,\mathrm{kHz}$ for our optical circuit. This represents an enhancement of roughly four orders of magnitude over our current implementation and can approach seven orders of magnitude ($\sim 1~\mathrm{MHz}$) for a lossless optical circuit. Alternatively, using the three-photon coincidence rate of \(148~\mathrm{kHz}\) reported by Cao \emph{et al.} \cite{SMWaltherGHZ}, we can estimate that the corresponding state-generation rate for our setup would be $\sim 0.1~\mathrm{kHz}$ and could be enhanced further, yielding an expected rate of \(37~\mathrm{kHz}\) for a lossless implementation of the unitary. Similarly, Maring \emph{et al.} \cite{SMquandelaGHZ} reported a three-photon on-chip coincidence rate exceeding \(5~\mathrm{kHz}\), which is also superior to that of our bulk-optical setup, and would yield a multi-mode NOON state generation rate on the order of 1~kHz, assuming a lossless realization of the unitary. However, despite their excellent multi-photon rates, even state-of-the-art quantum dot sources are not fully deterministic due to non-unit photon collection efficiencies. For example, Ding \emph{et al.}\cite{DingQD2025} reported a system efficiency of their single-photon source of 71\%. Cao \emph{et al.} \cite{SMWaltherGHZ} reported a fiber-coupled efficiency of 28.7\% for a bulk-optics-based single-photon source. Similarly, Maring \emph{et al.} \cite{SMquandelaGHZ} reported a 38.5\% source-to-fiber collection efficiency for a quantum dot source interfaced with an integrated photonic platform; this efficiency was further reduced to 9.13\% after propagation through the on-chip circuitry.
%\(55\%\) collection into the first lens and 70\% efficiency for fibre coupling, corresponding to source to fiber coupling efficiency of 38.5\%. 
Consequently, photon counting at the output remains necessary for quantum dot sources, even when a low-loss circuit is available.

\section{S4. Scalable generation of multi-mode two-photon NOON states}

Our scheme admits an extension that enables the generation of an arbitrary $d$-mode two-photon NOON state using $d$ single photons via a cascaded linear optical scheme.

\subsection{S4.1. Three-mode NOON state}

We gain insight into our extension procedure by noting that the first optical element of our implementation (a HWP) acts as a 50:50 beamsplitter between modes $A$ and $B$, as detailed in Sec.~S1. Acting on two separable single photons, this operation prepares an intermediate state $\ket{\psi^2_2}$ deterministically, namely a two-photon two-mode NOON state. 
\begin{equation}
\begin{split}
U_{\mathrm{BS1}}\ket{1,1,1,0}
&=
% \frac{1}{\sqrt{2}}\left(
% \ket{2,0,1,0}
% +
% \ket{0,2,1,0}
% \right)
\frac{1}{\sqrt{2}} \left(\ket{2,0} + \ket{0,2}\right)\otimes\ket{1,0} \\
&= \ket{\psi^2_2}\otimes\ket{1,0}
= \ket{\psi_{\mathrm{int}_3}}.
\end{split}
\end{equation}
The full protocol can then be broken down into two parts. The first part consists of producing the intermediate state $\ket{\psi_{\mathrm{int}_3}}$ through Hong-Ou-Mandel (HOM) interference in the first two modes, 
and the second part can be described by an effective four-mode unitary $U_{\mathrm{eff}}$ acting on this intermediate state, as shown in Fig~S5.
% 
% \begin{equation}
% \ket{In_{3}}
% =
% \frac{1}{\sqrt{2}} \left(\ket{2,0} + \ket{0,2}\right)\otimes\ket{1,0}.
% \end{equation}
%
The unitary reads
\begin{equation}
\textcolor{blue}{
U_{\mathrm{eff}}=
\begin{pmatrix}
1 & 0 & 0 & 0 \\[1mm]
0 & -\dfrac{1}{\sqrt{3}} & \sqrt{\dfrac{2}{3}} & 0 \\[1mm]
0 & -\dfrac{1}{\sqrt{3}} & -\dfrac{1}{\sqrt{6}} & -\dfrac{1}{\sqrt{2}} \\[1mm]
0 & -\dfrac{1}{\sqrt{3}} & -\dfrac{1}{\sqrt{6}} & \dfrac{1}{\sqrt{2}}
\end{pmatrix}.
}
\end{equation}
The complete transformation can be expressed as
\begin{equation}
U_{\mathrm{eff}}U_{\mathrm{BS1}}\ket{1,1,1,0}
=
\gamma_3
\ket{\psi_3^2}_{\mathrm{target}}
\ket{1}_{\mathrm{herald}}
+
\dots,
\end{equation}
with success probability $|\gamma_3|^2 = 1/4$.

\subsection{S4.2. Four-mode NOON state}

A four-mode two-photon NOON state can be defined as
\begin{equation}
\ket{\psi_4^2}
=
\frac{1}{2}
\left(
\ket{2,0,0,0}
+ e^{i\alpha_1}\ket{0,2,0,0}
+ e^{i\alpha_2}\ket{0,0,2,0}
+ e^{i\alpha_3}\ket{0,0,0,2}
\right).
\end{equation}

To generate this state with $\alpha_1=\pi, \alpha_2=0$ and $\alpha_3=0$, we enlarge the three-mode NOON state produced previously by introducing two additional modes: One containing a new single photon and one containing vacuum. The input state is
\begin{equation}
\ket{\psi_{\mathrm{int_4}}}
= \ket{\psi^2_3}\otimes\ket{1,0}
% \frac{1}{\sqrt{3}}
% \left(
% \ket{2,0,1,0,0}
% +
% \ket{0,2,1,0,0}
% +
% \ket{0,0,1,0,2}
% \right).
\end{equation}
The required five-mode unitary is
\begin{equation}
U_{5}
=
\begin{pmatrix}
1 & 0 \\
0 & U_{\mathrm{eff}}
\end{pmatrix},
\end{equation}
which acts as identity on the first mode and as $U_{\mathrm{eff}}$ on the second to fifth modes (see Fig.~S5). This yields
\begin{equation}
U_{5}\ket{\psi_{\mathrm{int}_4}}
=
\gamma_4
\ket{\psi_4^2}_{\mathrm{target}}
\ket{1}_{\mathrm{herald}}
+
\dots,
\end{equation}
with success probability $|\gamma_4|^2 = 4/18$, given that the 3-mode NOON state and single photon were available. Therefore, the total success probability of generating the 4-mode NOON state is given by the product $p_4 = |\gamma_3|^2\times|\gamma_4|^2 = 1/18$. 
\setcounter{figure}{4}
\begin{figure}
    \centering
    \captionsetup{labelfont={color=black}, textfont={color=black}}\includegraphics[width=0.75\linewidth]{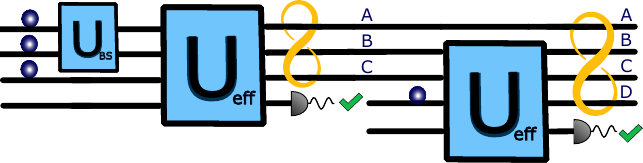} \caption{\textbf{Four-mode NOON state}: The scheme for extending a two-photon three-mode NOON state to four modes via a cascaded procedure.}  
    \label{fig: scalability}
\end{figure}

\subsection{S4.3. General $d$-mode NOON state}

The method can be iterated to generate a $d$-mode two-photon NOON state. We begin with a $(d-1)$-mode NOON state and introduce two additional modes: One containing a new single photon and one vacuum. The input state can be written as
\begin{equation}
\ket{\psi_{\mathrm{int}_d}}
=
% \frac{1}{\sqrt{d-1}}
% \sum_{k \in \mathcal{M}_d}
% \ket{0,\dots,0,\overset{k}{2},0,\dots,0}
\ket{\psi^2_{d-1}}
\otimes
\ket{1,0}.
%\qquad
%\mathcal{M}_d
%=
%\{1,\dots,d+2\}\setminus\{3,4\}.
\end{equation}

The transformation is implemented by a $(d+1)$-mode unitary
\begin{equation}
U_{d+1}
=
\begin{pmatrix}
I_{d-3} & 0 \\
0 & U_{\mathrm{eff}}
\end{pmatrix},
\end{equation}
which performs $U_{\mathrm{eff}}$ in the four-mode subspace and an identity transformation on the remaining $d-3$ modes. The output state takes the form
\begin{equation}
U_{d+1}\ket{\psi_{\mathrm{int}_d}}
=
\gamma_d
\ket{\psi_d^2}_{\mathrm{target}}
\ket{1}_{\mathrm{herald}}
+
\dots,
\end{equation}
with success probability
\begin{equation}
|\gamma_d|^2 = \frac{d}{6(d-1)}.
\end{equation}

After $d-2$ iterations, a $d$-mode two-photon NOON state is obtained using $d$ single photons and $2d-2$ modes in total. Of course, instead of verifying the intermediate states iteratively, the detection of all heralding photons can be delayed until the last iteration is complete. The overall success probability for $d \ge 3$ is given by
\begin{equation}
p_d
=
\frac{d}{2^{\,d-1} 3^{\,d-2}}.
\end{equation}

% For the same number of photon-number resources, 
This cascaded architecture \textcolor{blue}{requires fewer ancillary photons} than alternative heralded two-photon multi-mode NOON state schemes and yields a higher success probability for $d \le 7$~\cite{SMZhang2017}. 
% scheme of Zhang and Chan for two-photon multimode NOON states in $d \le 7$, reported in Ref. \cite{Zhang2017}. 
Moreover, any $U_{d+1}$ can be implemented using only two beam splitters. One of them is balanced while the other is unbalanced with $\theta= \arccos(-1/\sqrt{3})$ as explained in Sec.~S1.2. \textcolor{blue}{We note that for heralded schemes, the effective success probability can in principle be increased by multiplexing and is therefore not ultimately limited to the single-attempt success probability \cite{SMourreview}.}\\

\textcolor{blue}{\section{S5. Generation of multi-photon three-mode NOON states}}

\textcolor{blue}{Here we show that our scheme can be extended to increase the photon number of the multi-mode state. Specifically, this generalization enables the generation of arbitrary three-mode $(N+1)$-photon NOON states from three-mode $N$-photon NOON states by employing six ancillary photons, appropriate heralding measurements, and a cascaded linear optical network. We first present the protocol that transforms a two-photon NOON state into a three-photon state and then outline how to generalize the construction to the arbitrary-$N$ case.}

\textcolor{blue}{\subsection{S5.1. Three-photon three-mode NOON state}}

\textcolor{blue}{We begin from a heralded three-mode two-photon NOON state in three target modes $A,B,C$,
\begin{equation}
\ket{\psi^{2}_3}
=\frac{1}{\sqrt{3}}
\left(
\ket{2,0,0}
+e^{i\alpha_1}\ket{0,2,0}
+e^{i\alpha_2}\ket{0,0,2}
\right)_{ABC} \;.
\label{eq:input2}
\end{equation}
The goal is to increase its photon number in a heralded manner to
\begin{equation}
\ket{\psi^{3}_3}
=\frac{1}{\sqrt{3}}
\left(
\ket{3,0,0}
+e^{i\alpha_1}\ket{0,3,0}
+e^{i\alpha_2}\ket{0,0,3}
\right)_{ABC}\; ,
\label{eq:target3}
\end{equation}
using only passive linear optics, ancillary photons in Fock states, and photon-number-resolving detection.} 

\begin{figure}
    \centering
    \captionsetup{labelfont={color=blue}, textfont={color=blue}}\includegraphics[width=0.57\linewidth]{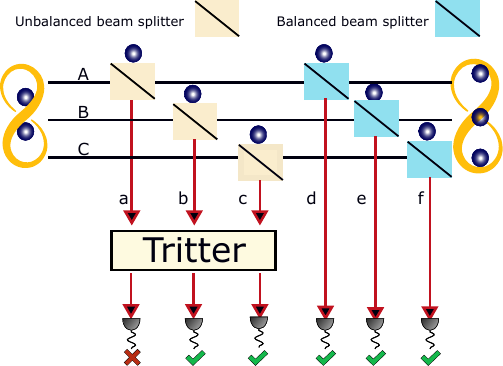} \caption{\textbf{Three-photon three-mode NOON state}: The scheme for extending a two-photon three-mode NOON state to three photons. A red cross denotes projection onto the vacuum state in mode $a$, while a green tick signals the detection of a single photon in that mode.}  
    \label{fig:higherN}
\end{figure}

\textcolor{blue}{This can be achieved by using a nine-mode unitary transformation 
\begin{equation}
U_{\mathrm{cat}}(\theta) =
\begin{pmatrix}
\dfrac{c}{\sqrt{2}}\mathbf{I}_3 & \dfrac{s}{\sqrt{2}}\mathbf{I}_3 & \dfrac{1}{\sqrt{2}}\mathbf{I}_3 \\[2mm]
-s\,\mathbf{F_3} & c\,\mathbf{F_3} & \mathbf{0_3} \\[2mm]
-\dfrac{c}{\sqrt{2}}\mathbf{I}_3 & -\dfrac{s}{\sqrt{2}}\mathbf{I}_3 & \dfrac{1}{\sqrt{2}}\mathbf{I}_3
\end{pmatrix},
\label{eq:Ucat}
\end{equation}
where $\mathbf{I}_3 $ and $\mathbf{0}_3$ are, respectively, the $3\times3$ identity and zero matrices, and
\begin{equation}
\mathbf{F_3}=
\frac{1}{\sqrt3}
\begin{pmatrix}
1 & 1 & 1\\
1 & \omega & \omega^2\\
1 & \omega^2 & \omega
\end{pmatrix},
\qquad
\omega=e^{2\pi i/3},
\qquad
z= \mathrm{cos}(\theta),
\qquad
s= \mathrm{sin}(\theta)\;.
\end{equation}
$U(\theta)$ is applied on the initial state:
\begin{equation}
\ket{\Psi_{\rm in}}
=
\ket{\psi^{2}_3}_{ABC}\ket{1,1,1,1,1,1}_{abcdef}\;.
\end{equation}
After the evolution $U(\theta)\ket{\Psi_{\rm in}}$, the detection of a single photon in each of the modes $b,c,d,e,f$ and vacuum in mode $a$ heralds the creation of a three-mode three-photon NOON state in the target modes $A,B$ and $C$:
\begin{equation}
  U(\theta)\cdot \ket{\Psi_{\rm in}}= \chi(\theta)\ket{\psi_3^3}_{ABC}\ket{0,1,1,1,1,1}_{abcdef}+... \; . 
\end{equation}
The state $\ket{\psi_3^3}_{ABC}$ can be heralded with a probability of $|\chi(\theta)|^2={\frac{(1-z^2)z^8}{48}}$. The transformation $\mathbf{F_3}$ is a standard three-mode Fourier transform, also known as a tritter. In Fig.~\ref{fig:higherN} we illustrate a linear optical realization of the circuit. It consists of three identical unbalanced beam splitters with transmittivity
$T=z^2=\cos^2(\theta)$, acting on mode pairs $(A,a)$, $(B,b)$, and $(C,c)$, followed by a tritter acting on modes $(a,b,c)$, and finally three balanced beam splitters acting on mode pairs $(A,d)$, $(B,e)$, and $(C,f)$.}

\textcolor{blue}{The newly generated $\ket{\psi_3^3}$ can now be used as a seed to generate $\ket{\psi_3^4}$ using the same unitary transformation, ancilla photons, modes, and heralding pattern. Cascading this circuit, one can in principle reach arbitrary $N$ with each step's success probability given by \begin{equation}
p_N(z_N) = K_N\,(1-z_N^2)\,z_N^{2N+4},
\qquad
K_N = \frac{N^2(N+1)}{144\cdot 2^N}.
\label{eq:pN_c}
\end{equation}}
\textcolor{blue}{Here, $z_{N}=\mathrm{cos(\theta_N)}= \frac{N+2}{N+3}$ maximizes the success probability. The overall success probability for generating an $N_f$-photon three-mode NOON state is given by
\begin{equation}
P_{\mathrm{total}}\big(3 \to N_f; \{z_N\}\big) \;=\; \prod_{N=3}^{N_f} p_N(z_N) \;.
\label{eq:Ptotal}
\end{equation}}

\textcolor{blue}{This three-mode protocol generalizes directly to an arbitrary number of modes, $d$. Given a prepared $N$-photon $d$-mode NOON state, one introduces two ancillary banks of $d$ modes, with each mode populated by one photon. 
% The resulting $3d$-mode network comprises $d$ identical tunable couplers between the signal and the first ancillary bank, a balanced $d$-mode Fourier interferometer acting on that bank, and $d$ final balanced couplers between the signal and the second ancillary bank. 
The resulting $3d$-mode network comprises $d$ identical unbalanced beam splitters between the output modes and the first ancillary bank, followed by a balanced $d$-mode Fourier interferometer acting on the ancillary bank. Subsequently, $d$ balanced beam splitters act on the alternate outputs of these unbalanced beam splitters and the second ancillary bank. 
Conditioning on one photon in every output of the second bank and a suitable $(d-1)$-photon record in the first bank produces, up to correctable phases, the target $(N+1)$-photon $d$-mode NOON state.}

\vspace{10pt}
\section{S6. Uncertainty estimation for the fidelity}
\label{supp:uncertainty_fidelity}
We now describe how statistical uncertainties in the experimentally measured quantities are propagated to the fidelity in Eq.~\eqref{eq:fidelity_optimized}, and, when relevant, to the corresponding optimal phases $\alpha_{1}^{\mathrm{opt}}$ and $\alpha_{2}^{\mathrm{opt}}$.
As mentioned in Sec.~S2.2., the fidelity estimator in Eq. \eqref{eq:supp:fidelity_full} depends on (i) the three-mode populations $P_{200}$, $P_{020}$, and $P_{002}$, and (ii) the pairwise coherence elements $\rho_{200,020}$, $\rho_{200,002}$, and $\rho_{020,002}$, which are inferred from the coincidence curves $C_{ij}(\theta)$ through the parameters $(A_{ij},V_{ij},\varphi_{ij})$ via Eqs.~\eqref{eq:supp:coefficients}.  
Because the mapping from measured counts to these quantities, and subsequently to the optimized fidelity in Eq.~\eqref{eq:fidelity_optimized}, is nonlinear, uncertainty propagation is carried out numerically using Monte Carlo sampling.
% This section details the statistical model used to assign error bars to the measured quantities entering the fidelity estimator and the procedure used to propagate these uncertainties to the reported fidelity and optimal phases.  The analysis proceeds in three steps: (i) estimation of pointwise uncertainties for normalized probabilities from shot-noise-limited four-fold counting statistics, (ii) weighted least-squares (WLS) fitting of interference fringes and calibration of the associated parameter covariance via the reduced chi-square, and (iii) Monte Carlo propagation of populations and fitted parameters through the nonlinear fidelity functional under physical consistency constraints.
\subsection{S6.1. Shot-noise model for experimentally obtained probabilities}
% \subsection{Shot-noise model for normalized four-fold outcome probabilities}
\label{sec:supp_shotnoise}

All probabilities entering the analysis are estimated from four-fold coincidence events.  
For a fixed measurement setting (e.g.\ a given wave-plate angle $\theta$ in a scan), we record detection events and retain only those in which one photon is detected in the heralding mode, one in the trigger mode, and corresponding coincidence clicks are obtained in the measurement modes.  
Each accepted event is then classified according to the photon-number distribution in the two modes under consideration after the projection step (see Sec.~\ref{sec:supp:MeasurementStrategy}).  
In the two-mode measurements, this classification corresponds to the outcomes $(1,1)$, $(2,0)$, and $(0,2)$, referring exclusively to the photon-number distribution in the measured pair of modes. 
We therefore label the possible outcomes by an index $k \in \{(1,1),(2,0),(0,2)\}$, and denote by $n_k$ the number of events registered for each outcome $k$, with 
\begin{equation}
    N=\sum_k n_k
\end{equation}
the total number of four-fold events acquired for that setting, where the corresponding normalized counts are
\begin{equation}
    p_k=\frac{n_k}{N}.
\end{equation}
Assuming shot-noise-limited counting statistics, the raw counts may be modeled as independent Poisson variables. 
% $n_k\sim\mathrm{Pois}(\lambda_k)$. 
Conditioning on $N$ leads to a multinomial model, from which the variances of the normalized probabilities follow, to leading order in $1/N$, as
\begin{equation}
\mathrm{Var}(p_k) \simeq \frac{p_k(1-p_k)}{N}.
\label{eq:supp:shotnoise}
\end{equation}
% 
% Conditioning on the observed total $N$ then yields the multinomial model
% % 
% \begin{equation}
%     (n_1,n_2,\dots)\,|\,N \sim \mathrm{Multinomial}\!\left(N;\,\pi_1,\pi_2,\dots\right),
%     \qquad \sum_k\pi_k=1.
% \end{equation}
% % 

% 
For each reduced two-mode interference scan, the coincidence probability $C_{ij}(\theta)$ is evaluated at discrete angles $\{\theta_m\}_{m=1}^M$ from a set of mutually exclusive outcomes [e.g.\ $(11),(20),(02)$], with $N_m=n_{11}(\theta_m)+n_{20}(\theta_m)+n_{02}(\theta_m)$ and $p_{11}(\theta_m)=n_{11}(\theta_m)/N_m$.
The associated variance is then obtained directly from Eq.~\eqref{eq:supp:shotnoise} as
\begin{align}\label{eq:supp:sigma}
\sigma_m^2 \simeq \frac{p_{11}(\theta_m)\left[1-p_{11}(\theta_m)\right]}{N_m}.
\end{align}
Similarly, the populations $P_{200},P_{020},P_{002}$ entering Eq.~\eqref{eq:supp:fidelity_full} are obtained as normalized counts from the same underlying four-fold outcome register.  In  the multinomial description, these estimated probabilities are not independent but exhibit anticorrelations due to the shared normalization constraint.
However, in the Monte Carlo propagation below we approximate $P_{200},P_{020},P_{002}$ as independent Gaussian variables with the experimentally determined standard errors. 
For the event numbers in our experiment, this approximation has a negligible effect on the final uncertainty, which is dominated by the fringe-fit contributions, while significantly simplifying the sampling procedure. 
 % (A fully consistent multinomial/Dirichlet sampling of the populations yields compatible results within the reported error bars.)

\subsection{S6.2. Extraction of fringe parameters and their uncertainties}

Each coincidence curve $C_{ij}(\theta)$ is fit to the sinusoidal form introduced in Eq.~\eqref{eq:supp:coincidence_form}.  
For numerical convenience, we use the equivalent linear parametrization
\begin{equation}
C_{ij}(\theta) = a_0 + A_c \cos(8\theta) + A_s \sin(8\theta),
\label{eq:supp:linear_model}
\end{equation}
from which the physical parameters are obtained as
\begin{equation}
A_{ij} = 2a_0,\qquad
V_{ij} = 2\sqrt{A_c^2 + A_s^2},\qquad
\varphi_{ij} = -\mathrm{atan2}(A_s,A_c),
\label{eq:supp:AVphi_def}
\end{equation}
where $\mathrm{atan2}(A_s,A_c)$ denotes the two-argument arctangent,  i.e.\ the phase of $A_c+iA_s$, taken in the principal interval $(-\pi,\pi]$.
This parametrization is fully equivalent to Eq.~\eqref{eq:supp:coincidence_form}, but is more convenient for the weighted least-squares analysis described below.
The fit is performed using weighted least squares (WLS), with weights determined by the shot-noise uncertainties of each point. 
For a given scan, let $y=(y_1,\dots,y_M)^{\mathsf T}$ denote the measured coincidence probabilities at angles $\{\theta_m\}_{m=1}^M$ [$y_m \equiv p_{11}(\theta_m)$], and let $\beta=(a_0,A_c,A_s)^{\mathsf T}$ denote the fit parameters.
The corresponding design matrix is
\begin{equation}
X \equiv
\begin{pmatrix}
1 & \cos(8\theta_1) & \sin(8\theta_1)\\
\vdots & \vdots & \vdots \\
1 & \cos(8\theta_M) & \sin(8\theta_M)
\end{pmatrix},
\end{equation}
and the pointwise variances obtained in Eq.~\eqref{eq:supp:sigma} define the diagonal covariance matrix
\begin{equation}
\Sigma \equiv \mathrm{diag}(\sigma_1^2,\dots,\sigma_M^2).
\end{equation}
The WLS estimator is then obtained by minimizing
\begin{align}
\chi^2(\beta) = (y-X\beta)^{\mathsf T}\, \Sigma^{-1}\, (y-X\beta),
\label{eq:chi2_def}
\end{align}
where we denote the minimizing parameter vector by $\beta^*$.
Since bounds are imposed in the optimization (specifically, $a_0\in[0,1/2]$), the solution is obtained numerically.

If the model in Eq.~\eqref{eq:supp:linear_model} correctly describes the data and the variances $\sigma_m^2$ fully capture the fluctuations of the measured points, then the minimized statistic
\begin{equation}
\chi^2_{\min}\equiv \chi^2(\beta^*)
\end{equation}
is expected to satisfy
\begin{equation}
\frac{\chi^2_{\min}}{\nu}\approx 1,
\qquad
\nu=M-p,
\qquad
p=3,
\end{equation}
where $\nu$ is the number of degrees of freedom and $p$ is the number of fitted parameters.
% 
% Under the assumptions that (i) the linear model in Eq.~\eqref{eq:supp:linear_model} is correct, (ii) the pointwise uncertainties $\sigma_m$ fully capture the variance of $C_{ij}(\theta_m)$, and (iii) residuals are independent and approximately Gaussian (appropriate for large count numbers), the statistic
% \begin{align}
% \chi^2_{\min} \equiv \chi^2(\check{\beta})
% \end{align}
% is distributed as a chi-square random variable with
% \begin{align}
% \nu = N - p,\qquad p=3
% \end{align}
% degrees of freedom, and therefore $\mathbb{E}[\chi^2_{\min}/\nu]\approx 1$. 
%

% TODO: I should say what was $\chi^2_{\min}/\nu$ for the reported data
% 
Values $\chi^2_{\min}/\nu>1$ indicate that the observed point-to-point fluctuations are larger than expected from the shot-noise model alone.
For the three scans considered here, we obtain $\chi^2_{\min}/\nu = 2.26,\,1.16,\,3.77$, indicating that two scans exhibit fluctuations larger than expected from the shot-noise model, while one is consistent with it within statistical uncertainty.
In sequential phase scans, this may arise from additional experimental fluctuations and/or correlations between neighboring scan points, such as slow drifts during data acquisition. Rather than attributing this excess variability to a specific mechanism, we incorporate it conservatively as a calibration of the fit uncertainty.
Accordingly, the covariance matrix of the fitted parameters is taken to be
\begin{equation}
\mathrm{Cov}(\beta^*)
=
\frac{\chi^2_{\min}}{\nu}\,
(X^{\mathsf T}\, \Sigma^{-1}\, X)^{-1}.
\label{eq:supp:cov_beta}
\end{equation}
Equivalently, this corresponds to replacing the shot-noise variances $\sigma_m^2$ by effective variances $(\chi^2_{\min}/\nu)\, \sigma_m^2$, thereby preserving the relative weighting of the data points while matching the overall level of fluctuations observed in the scan.
% 
% We treat $\check{\sigma}^2_{\mathrm{eff}}>1$ as \emph{overdispersion} and incorporate it as a conservative calibration of the fit covariance, without attributing it to a specific physical mechanism.

% 
Finally, the covariance of the physical parameters $(A_{ij},V_{ij},\varphi_{ij})$ is obtained from Eq.~\eqref{eq:supp:cov_beta} by standard first-order error propagation applied to the transformation in Eq.~\eqref{eq:supp:AVphi_def}. These covariances are then used in the Monte Carlo propagation of the fidelity uncertainty described in the next subsection.

% \paragraph{Transformation to physical fringe parameters.}
% The fitted coefficients are mapped to the physically reported parameters
% \begin{align}
% A = 2a_0,\qquad V = 2\sqrt{A_c^2+A_s^2},\qquad \phi = -\mathrm{atan}2(A_s,A_c),
% \label{eq:AVphi_def}
% \end{align}
% where $A$ is an offset, $V$ is the visibility amplitude (prior to imposing physical bounds), and $\phi$ is the fitted phase.  Let $g:\beta\mapsto(A,V,\phi)$ denote this mapping.  The covariance of $(A,V,\phi)$ is obtained by first-order (delta-method) propagation:
% \begin{align}
% \mathrm{Cov}(A,V,\phi) \;=\; J\, \mathrm{Cov}(\check{\beta})\, J^{\mathsf T},
% \qquad
% J \equiv \left.\frac{\partial(A,V,\phi)}{\partial(a_0,A_c,A_s)}\right|_{\check{\beta}}.
% \label{eq:cov_AVphi}
% \end{align}
% Explicitly, writing $r=\sqrt{A_c^2+A_s^2}$,
% \begin{align}
% J =
% \begin{pmatrix}
% 2 & 0 & 0\\
% 0 & \dfrac{2A_c}{r} & \dfrac{2A_s}{r}\\[6pt]
% 0 & -\dfrac{A_s}{r^2} & \dfrac{A_c}{r^2}
% \end{pmatrix}_{\check{\beta}}.
% \label{eq:jacobian_AVphi}
% \end{align}
% When reporting visibilities we enforce the physically allowed range via clipping, $V\mapsto \min\{1,\max\{0,V\}\}$.  This clipping is used only for reporting and for enforcing physical constraints during sampling; the linear covariance propagation \eqref{eq:cov_AVphi} remains valid in the neighborhood of $\check{\beta}$ when the estimate is not saturated at the boundary.

\subsection{S6.3. Monte Carlo propagation to fidelity}

We now propagate the uncertainties of the measured quantities to the fidelity defined in Eq.~\eqref{eq:fidelity_optimized}. 
Since the fidelity depends nonlinearly on the measured populations and fitted fringe parameters, and involves an internal optimization over phase variables, analytic error propagation is not practical.
We therefore perform this propagation using Monte Carlo sampling.
For each Monte Carlo realization $k$, we generate a set of sampled parameters as follows:
\begin{itemize}
\item Populations $P_{200}^{(k)}, P_{020}^{(k)}, P_{002}^{(k)}$  are drawn from Gaussian distributions centered at their measured values.
\item Fringe parameters $(A_{ij}^{(k)},V_{ij}^{(k)},\varphi_{ij}^{(k)})$ for each mode pair $(ij)\in\{AB,AC,BC\}$ are drawn from multivariate normal distributions defined by their fitted covariance matrices.
\end{itemize}

Each sampled set is required to satisfy the physical constraints
\begin{equation}
% 0 \le V_{ij}^{(k)} \le A_{ij}^{(k)}, \qquad
|\rho_{mn}^{(k)}|^2 \le \rho_{mm}^{(k)}\rho_{nn}^{(k)},
\end{equation}
% consistent with Eqs.~\eqref{eq:supp:constraints}
which express positivity of the underlying density matrix and are implemented via the equivalent constraints on observable quantities given in Eqs.~\eqref{eq:supp:constraints}.
Samples that violate these constraints are discarded.
For each accepted sample, the fidelity function $F^{(k)}(\alpha_1,\alpha_2)$ is evaluated using Eq.~\eqref{eq:supp:fidelity_observables}.  
The optimized fidelity is then obtained as
\begin{equation}
F^{(k)} = \max_{\alpha_1,\alpha_2} F^{(k)}(\alpha_1,\alpha_2),
\end{equation}
with the corresponding maximizing phases $(\alpha_1^{\mathrm{opt}\, (k)},\alpha_2^{\mathrm{opt}\, (k)})$.
% in direct analogy with Eq.~\eqref{eq:fidelity_optimized}.  

From the ensemble of accepted samples $\{F^{(k)},\alpha_1^{\mathrm{opt}\, (k)},\alpha_2^{\mathrm{opt}\, (k)}\}_{k=1}^{K}$, we report the fidelity and phase estimates as sample means, with uncertainties given by the corresponding standard deviations,
\begin{equation}
\overline{F} = \frac{1}{K}\sum_{k=1}^{K}F^{(k)},\qquad
\sigma_F = \sqrt{\frac{1}{K-1}\sum_{k=1}^{K}\big(F^{(k)} - \overline{F}\big)^2},
\end{equation}
and analogously for $\alpha_1$ and $\alpha_2$.

% This procedure consistently propagates the combined effects of counting statistics, fitting uncertainty in the fringe parameters, and the nonlinear optimization in Eq.~\eqref{eq:fidelity_optimized} to the reported fidelity.
This procedure accounts for the combined effect of shot noise, fitting uncertainty in the fringe parameters, and the nonlinear optimization in Eq.~\eqref{eq:fidelity_optimized}, providing a reliable estimate of the uncertainty associated with the reported fidelity.

% ============================================================

\renewcommand{\refname}{Supplemental References}

\end{document}